\documentclass[superscriptaddress,reprint,aps,prb]{revtex4-1}

\usepackage{graphicx}
\usepackage{dcolumn}
\usepackage{bm}

\begin{document}

\title{Experimental and Theoretical Study of Electronic and Hyperfine Properties of Hydrogenated Anatase (TiO$_2$): Defects Interplay and Thermal Stability}

\author{D. V. Zyabkin}
\email{dmitry.zyabkin@tu-ilmenau.de}
\affiliation{Chair Materials for Electrical Engineering and Electronics, Institute of Materials Science and Engineering, Institute of Micro and Nanotechnologies MacroNano\textregistered, TU Ilmenau, Gustav-Kirchhoff-Strasse 5, 98693 Ilmenau, Germany}
\affiliation{Department of Experimental Physics, Faculty of Science, Palack{\'y} University in Olomouc, 17. Listopadu 12, 77147 Olomouc, Czechia}
\author{H. P. Gunnlaugsson}
\affiliation{Science Institute, University of Iceland, Dunhaga 3, 107 Reykjav\'{i}k, Iceland}
\author{J. N. Gon\c{c}alves}
\affiliation{Departmento de F\'{i}sica and CICECO, Universidade de Aveiro, 3810-193 Aveiro, Portugal}
\author{K. Bharuth-Ram}
\affiliation{School of Chemistry \& Physics, University of KwaZulu-Natal, Durban 4000, South Africa}
\affiliation{Physics Department, Durban University of Technology, Durban 4000, South Africa}
\author{B. Qi}
\affiliation{Science Institute, University of Iceland, Dunhaga 3, 107 Reykjav\'{i}k, Iceland}
\author{I. Unzueta}
\affiliation{Department of Applied Mathematics, University of the Basque Country (UPV/EHU), Torres Quevedo Ingeniaria Plaza 1, 48013 Bilbao, Spain}
\author{D. Naidoo}
\affiliation{School of Physics, University of the Witwatersrand, Johannesburg, 2050, South Africa}
\author{R. Mantovan}
\affiliation{CNR-IMM, Unit of Agrate Brianza, Via Olivetti 2, 20864 Agrate Brianza (MB), Italy}
\author{H. Masenda}
\affiliation{School of Physics, University of the Witwatersrand, Johannesburg, 2050, South Africa}
\author{S. {\'O}lafsson}
\affiliation{Science Institute, University of Iceland, Dunhaga 3, 107 Reykjav\'{i}k, Iceland}
\author{G. Peters}
\affiliation{School of Physics, University of the Witwatersrand, Johannesburg, 2050, South Africa}
\author{J. Schell}
\affiliation{European Organization for Nuclear Research (CERN), CH-1211 Geneva, Switzerland}
\affiliation{Institute for Materials Science and Center for Nanointegration Duisburg-Essen (CENIDE), University of Duisburg-Essen, 45141 Essen, Germany}
\author{U. Vetter}
\affiliation{Chair Materials for Electrical Engineering and Electronics, Institute of Materials Science and Engineering, Institute of Micro and Nanotechnologies MacroNano\textregistered, TU Ilmenau, Gustav-Kirchhoff-Strasse 5, 98693 Ilmenau, Germany}
\author{A. Dimitrova}
\affiliation{Institut f{\"u}r Physik and Institut f{\"u}r Mikro- und Nanotechnologien, Technische Universit{\"a}t Ilmenau, PF 100565, 98684 Ilmenau, Germany}
\author{S. Krischok}
\affiliation{Institut f{\"u}r Physik and Institut f{\"u}r Mikro- und Nanotechnologien, Technische Universit{\"a}t Ilmenau, PF 100565, 98684 Ilmenau, Germany}
\author{P. Schaaf}
\affiliation{Chair Materials for Electrical Engineering and Electronics, Institute of Materials Science and Engineering, Institute of Micro and Nanotechnologies MacroNano\textregistered, TU Ilmenau, Gustav-Kirchhoff-Strasse 5, 98693 Ilmenau, Germany}

\begin{abstract}
    Performance of TiO$_2$-based materials is highly dependent on the electronic structure and local defect configurations. Hence, a thorough understanding of defects interaction plays a key role. In this study we report on the results from emission $^{57}$Fe M{\"o}ssbauer Spectroscopy experiments, using dilute $^{57}$Mn implantation into pristine (TiO$_2$) and hydrogenated anatase held at temperatures between 300-700 K. Results of the electronic structure and local environment are complemented with \textit{ab-initio} calculations. Upon implantation both Fe$^{2+}$ and Fe$^{3+}$ are observed in pristine anatase, where the latter demonstrates the spin-lattice relaxation. The spectra obtained for hydrogenated anatase show no Fe$^{3+}$ contribution, suggesting that hydrogen acts as a donor. Due to the low threshold, hydrogen diffuses out of the lattice. Thus showing a dynamic behaviour on the time scale of the $^{57}$Fe 14.4 keV state. The surrounding oxygen vacancies favor the high-spin Fe$^{2+}$ state. The sample treated at room temperature shows two distinct processes of hydrogen motion. The motion commences with the interstitial hydrogen, followed by switching to the covalently bound state. Hydrogen out-diffusion is hindered by bulk defects, which could cause both processes to overlap. Supplementary UV-Vis and electrical conductivity measurements show an improved electrical conductivity and higher optical absorption after the hydrogenation. X-ray photoelectron spectroscopy at room temperature reveals that the sample hydrogenated at 573 K shows presence of both Ti$^{3+}$ and Ti$^{2+}$ states. This could imply that a significant amount of oxygen vacancies and -OH bonds are present in the samples. Theory suggests that in the anatase sample implanted with Mn(Fe), probes were located near equatorial vacancies as next-nearest-neighbours, whilst a metastable hydrogen configuration is responsible for the annealing behaviour. The obtained information provides a deep insight into elusive hydrogen defects and their thermal stability.
\end{abstract}

\maketitle 

\section{INTRODUCTION}
Recently, anatase has attracted significant attention due to good performance for water splitting with high yields under ultraviolet exposure. High photocatalytic activity, in contrast to other TiO$_2$ polymorphs (rutile and brookite), is attributed to lofty carrier separation efficiency, mainly due to a different octahedral arrangement of TiO$_{6}$~\cite{Stevanovic2013,Stevanovic2012}. Furthermore, the material is abundant, stable and highly reactive~\cite{Kurian2013,Lu2011,Hoffmann1995}. Recently, the reduction of anatase in different atmospheres (e.g. hydrogen) has triggered numerous experiments towards its application (from hydrogen production to photocatalytic reduction). The approach is based on the principle of introducing a plethora of defects, creating enough lattice disorder at the surface or near surface regions of TiO\textsubscript{2} (i.e. oxygen deficiency and hydrogen doping)~\cite{Chen2011,Xia2013,HWang2019}. During hydrogenation, atomic hydrogen impinges on the material surface and can get absorbed spontaneously into oxygen or titanium sites. Upon absorption it could diffuse within the sub-surface layers of TiO\textsubscript{2} and may lead to generation of oxygen vacancies via water desorption and hydrogen doping~\cite{Wang2019,Syzgantseva2011,Aschauer2012}. Such treatments have shown to remarkably intensify both visible and infrared absorption (IR), thereby increasing the yields of hydrogen production above most photocatalysts of this type. The structure of hydrogenated anatase is identified as consisting of a crystalline lattice core surrounded by an amorphous shell (such as  Ti$_{4}$O$_{7}$) containing the hydrogen dopant. The exposed layer is in order of 1-2 nm, although hydrogen easily diffuses into bulk~\cite{Xia2013,Chen2015,Chen2011}. Besides the intentional lattice dopant, hydrogen is well-known to be a common impurity in semiconductors with a prominent impact on electronic and optical properties. Theoretical investigations have shown that hydrogen in TiO$_2$ primarily tends to bond with oxygen and then acts as a donor or as an amphoteric  defect~\cite{Xiong2007,Li2014}.

Even though there is a lack of experimental data available on anatase doped with hydrogen,  the results obtained on rutile vary remarkably. Recent studies on rutile have shown that the electron associated with a hydrogen centre is not delocalised (same as anatase) as one may assume for a shallow effective mass donor, but localised close to a Ti ion, as a polaron~\cite{Brant2011,Sezen2014}. There are few IR absorption studies on hydrogenated anatase which show that after hydrogenation several new lines become evident. This has been attributed to the stretch of local vibrational mode of -OH bonds~\cite{Lavrov2015,Lavrov2016}. 

The photocatalytic efficiency of TiO$_2$-based materials is not only confined by the large band gap (3.2 keV for anatase), but also by the losses due to the defect-induced recombination~\cite{Yu2013}. On the contrary, when defects such as oxygen vacancies (V$_o$) or Ti$^{3+}$ are induced in small quantities, they affect the photocatalytic performance positively~\cite{Chen2015}. Therefore, it is of great importance to understand the role of defects and their interactions in hydrogenated TiO$_2$.

In the present work, we report on studies of the defect and electronic structure of pristine and hydrogenated anatase TiO$_2$ thin films by means of $^{57}$Fe emission M{\"o}ssbauer spectroscopy (eMS)~\cite{Johnston2017} following the implantation of $^{57}$Mn ($t_{1/2}$=1.5 min.) in a temperature range from $\sim$ 295 to 700 K. This method is one of the most powerful, which allows to precisely probe the local environment and detect subtle changes without its alteration. In addition, we complement the eMS studies with X-ray photoelectron spectroscopy (XPS), resistivity measurements, grazing incidence X-ray diffraction (GIXRD) and \textit{ab-initio} density functional theory calculations within \textsc{vasp} and \textsc{wien2k} environments~\cite{KRESSE1996,blaha2008}. 

In order to develop a clear picture of hydrogen behaviour in thin films, two distinctively hydrogenated samples have been studied. The obtained results show that hydrogen has strong influence on the electronic structure (in comparison to pristine samples) and shows a behaviour depending on the hydrogenation degree. We could clearly observe two types of hydrogen present in the samples - interstitial and covalently bonded. Oxygen vacancies may potentially support hydrogen in the lattice up to high temperatures. Theory suggests that the defects interplay is more complex than previously expected. 
	
   \section{EXPERIMENTAL DETAILS}

  \subsection{Sample preparation}

    TiO$_2$ thin films were deposited onto Si and quartz substrates by radio frequency (RF) sputtering (LA 440S by VON ARDENNE) utilizing a ceramic TiO$_2$ target (99.9\% FHR Anlagenbau, Germany). During the process, the sputtering power was maintained at 200 W and the Ar flux was fixed at 80 sccm. The thickness of thin films was measured with an ellipsometer (Sentech SE 801) and the crystal structure was determined by virtue of GIXRD. Prior to hydrogenation, the samples were annealed at 773 K for 3 hours to ensure their crystallization into anatase. Thereafter, the samples were treated in a chamber for plasma-enhanced hydrogenation treatment, where an inductively coupled plasma (ICP) instrument (Plasma Lab 100 ICP-CVD, Oxford Instruments) was used. The H$_{2}$ plasma treatment was performed for 30 minutes at  two distinct temperatures: room (RT, 293 K) and 573 K, producing the TiO$_2$:H-RT and TiO$_2$:H-573 samples, respectively. The ICP power was held at 2000 W, the chamber pressure at 3.44-3.6 Pa, and the H$_{2}$ flow rate at 50 sccm.

    \subsection{Sample characterization}
    The samples were characterised by GIXRD, UV-Vis optical absorption, XPS and resistivity measurements, while information on the electronic structure during annealing was obtained from emission M{\"o}ssbauer Spectroscopy experiments.

    GIXRD patterns were recorded with a SIEMENS/BRUKER D5000 X-ray diffractometer using Cu-K$\alpha$ radiation at 40 kV and 40 mA, with the samples being scanned from $23^\circ$ to $80^\circ$ with a step size of $0.02^\circ$. Optical absorption in the UV to visible wavelength range was measured using a UV-Vis spectrometer (Varian Cary 5000 UV-Vis-NIR). Room temperature (RT) resistivity measurements were conducted using a 10 V four-point probe setup (Jandel engineering) in linear Van der Pauw arrangement. XPS analysis was performed utilizing the normal emission utilizing monochromatic Al-K$\alpha$ (hv = 1486.7 eV) radiation. Further details of the experimental setup are described in Ref.~\cite{Vlaic2015}. Core level spectra were measured at constant pass energy (13 eV) with a total energy resolution of 0.6 eV in absence of charge neutralization as well as a further binding energy correction. The correction of the binding energy was made with the C 1s peak, assigned with a binding energy of 285 eV. The accuracy for binding energy assignment is estimated to be $\pm$0.2 eV.

    $^{57}$Fe eMS measurements were carried out at the ISOLDE/ CERN facility~\cite{Borge2017,Catherall2017}, where the parent radioactive isotope was produced with 1.4 GeV proton induced fission in a heated UC$_{2}$ target and further element-selective laser ionization~\cite{Fedoseyev2000}. Following acceleration to 40 keV and magnetic mass separation, the $^{57}$Mn ions were implanted into the polycrystalline anatase TiO$_2$ samples held at the implantation/measurement temperature ranging from 295 to 735 K within an implantation chamber. Measurements were performed with different temperature steps, in order to check a possible presence of other transformations. Samples were mounted with surface normal at $30^\circ$ relative to the beam direction. Emission M{\"o}ssbauer spectra were recorded during the implantation using a resonance detector equipped with a $^{57}$Fe enriched stainless steel electrode, mounted on a conventional drive system outside the implantation chamber at $60^\circ$ relative to the sample's normal.

    In all cases, the samples were implanted with a total fluence (5-6 measurements per sample) below 3 $\times$ 10$^{12}$ $^{57}$Mn ions/cm$^{2}$ thus providing a truly dilute concentration. From SRIM calculations~\cite{Ziegler2010}, the ion range was estimated to be 22 nm, the maximum local concentration being~$\sim$ 10$^{-3}$ at.\%. Isomer shift (IS) values are given relative to $\alpha$-Fe at RT.

   \section{RESULTS AND DISCUSSION}

    \subsection{\label{sec:level5}General characterization}

\begin{figure}[ht]
            \centering
            \includegraphics[width=9cm, trim=0.15cm 0.4cm 0cm 0cm, clip]{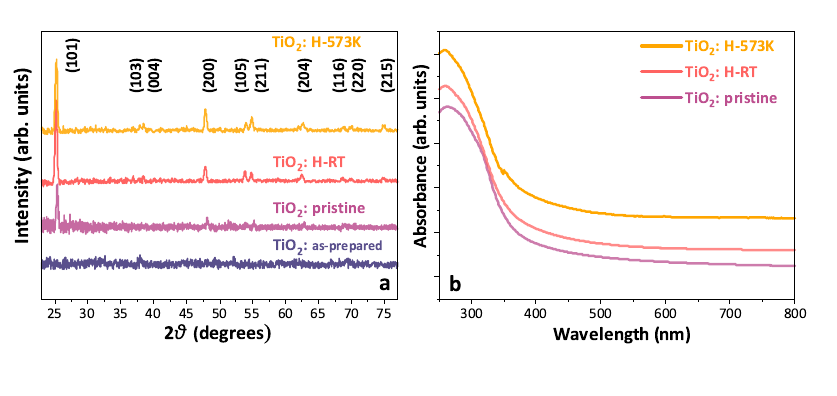}
            \caption{(a) XRD pattern and (b) UV-Vis absorption spectra of as-prepared TiO$_2$ and after annealing (pristine), the samples hydrogenated at RT (TiO$_2$:H-RT) and the sample hydrogenated at 573 K (TiO$_2$:H-573K).}
            \label{fig:XRDUV}
            \end{figure}

    Fig.~\ref{fig:XRDUV}(a) shows the GIXRD patterns of the anatase thin films, as prepared (TiO$_2$:as-prepared), after 3 hours of annealing at 773 K (TiO$_2$: pristine) and the two hydrogen treated samples (TiO$_2$: H-RT and TiO$_2$: H-573K). The films were 100(7) nm thick. The results show that the as-prepared film is amorphous, which, after 773 K annealing, is crystallised into the single anatase phase, with the diffraction reflections corresponding close to expectations (PDF-89-4921). After hydrogenation, the diffraction peaks increased slightly in intensity and became more clear. Neither rutile phase nor TiH$_{2}$ is observed in the GIXRD patterns.

    After the hydrogenation, the samples changed their colour from white (pristine) to grey (TiO2:H-RT) and black (TiO2:H-573) reflecting changes in their optical and structural properties.
    Sun \textit{et al.}~\cite{Sun2011} have reported that the amount of hydrogen stored in the TiO$_2$ crystals may be dependent on their facets, which in turn could also affect the colour changes (higher hydrogen concentration equals to darker colour). The amount of stored hydrogen was around 1.4 wt.\% on the (101) surface (which is the most stable facet~\cite{Diebold2003}), whilst 1.0 wt.\% was found on the (001) surface. These authors commented that hydrogen should occupy the interstitial sites in the titanium-oxygen octahedra~\cite{Sun2011}. The effects of hydrogenation on the UV-Vis absorption spectra of the films are presented in Fig.~\ref{fig:XRDUV}(b). The impact of hydrogenation at both temperatures is visible. The optical response of hydrogenated TiO$_2$ shows a clear increase in the absorbance, which is the characteristic of the oxygen vacancy impact on the light absorption properties of TiO$_2$~\cite{Naldoni2012}. The TiO2:H-573 sample extends the absorption edge of TiO$_2$ in the UV-visible range with a small red shift. The band gap can be calculated using the Tauc model, where the band gap is dependent on equivalent absorption coefficient and the energy of photons~\cite{Tauc2006}. From these calculations, the sample hydrogenated at room temperature exhibits decreased band gap by 0.04 eV, while the 573 K sample shows a reduced band gap by 0.13 eV. These changes are similar to the finding of previous works~\cite{He2013,Mohammadizadeh2015}. Therefore, the UV-visible absorption spectra indicate the creation of oxygen vacancies in TiO$_2$ after hydrogenation.
            
     \begin{figure*}
                 \centering
                 \includegraphics[width=16.5cm, trim=0.5cm 1cm 0.5cm 1cm, clip]{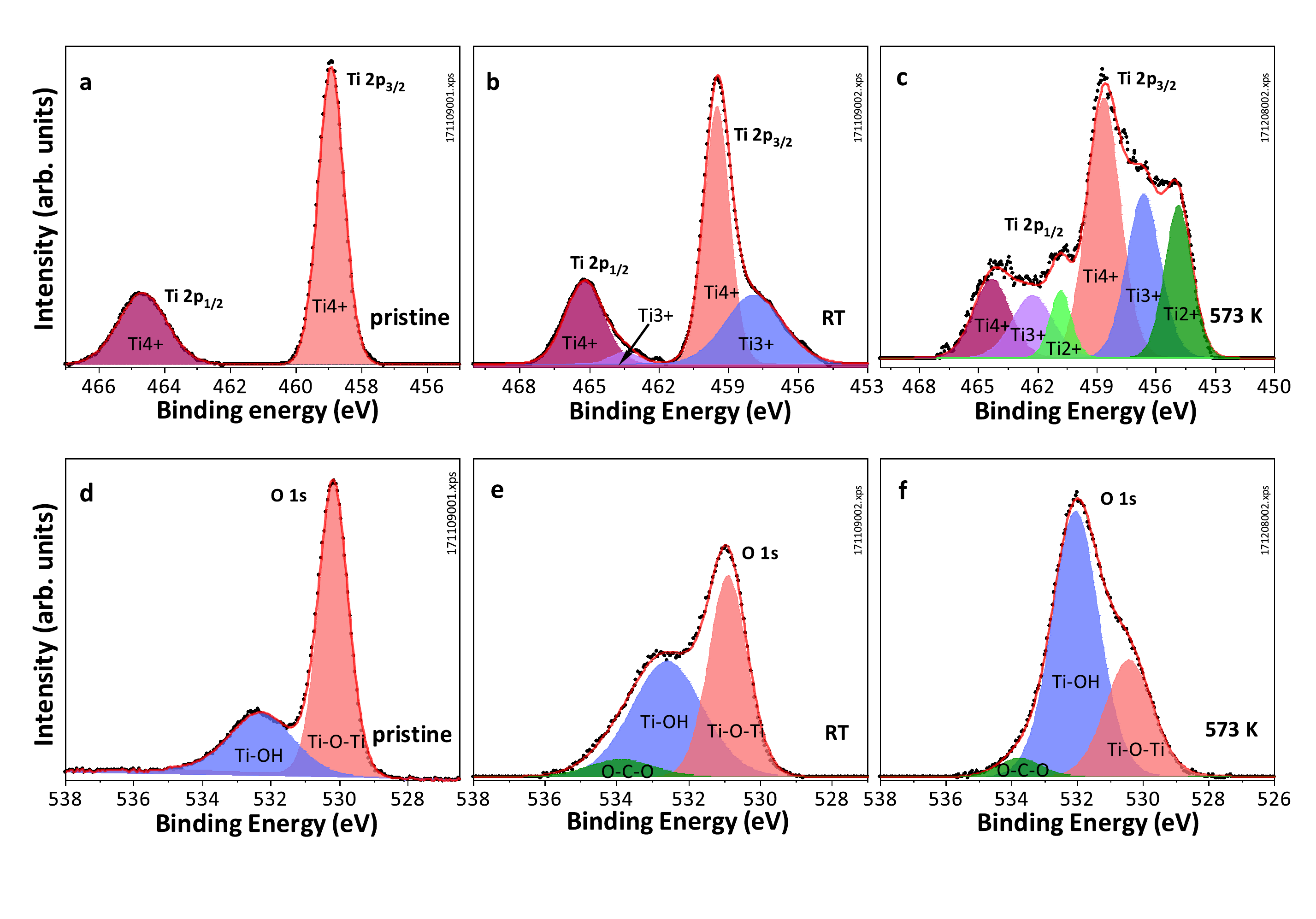}
                 \caption{XPS spectra for Ti 2p (a-c) and O 1s (d-f) of TiO$_2$ films before (a,d) and after hydrogenation treatments performed at room (b,e) and 573 K temperature (c,f).}
                 \label{fig:XPS}
            \end{figure*}

    Resistivity measurements show the gradual decrease of resistivity with the hydrogenation treatments. The pristine TiO$_2$ films showed a resistivity greater than 2 $\times$ 10$^{5}$ $\Omega$$\cdot$m, which is above the instrumentation limitation. However, the sample hydrogenated at RT shows a rapid decrease to 6.2 $\times$ 10$^{-6}$ $\Omega$$\cdot$m while the sample treated at 573 K shows a further decrease to 4.3 $\times$ 10$^{-6}$ $\Omega$$\cdot$m. This can be interpreted as the release of electrons trapped in V$_o$ or deficient Magn{\'e}li formations~\cite{Noerenberg1998}. Therefore, in correlation with optical absorption results, the higher V$_o$ concentration, the greater the electron participation in conductivity~\cite{Cronemeyer1959}.

    The near-/surface chemical bonding and electronic valence band positions of the TiO$_2$ films were investigated with XPS. Ti 2p and O 1s core level spectra were measured in order to evaluate the chemical states of Ti and O in TiO$_2$. The spectra of Ti 2p for the TiO$_2$ film before and after hydrogenation under different conditions are shown in Figs.~\ref{fig:XPS}(a-c). For the pristine sample shown in Fig.~\ref{fig:XPS}(a), only peaks due to Ti$^{4+}$ are observed at 458.9 eV (2p$_{3/2}$) and 464.6 eV (2p$_{1/2}$), which are well-known binding energies in pristine TiO$_2$~\cite{Ivanov2016}. The spectrum for the sample hydrogenated at RT depicted in Fig.~\ref{fig:XPS}(b) shows evidence of reduction of Ti$^{4+}$ ions to Ti$^{3+}$ with two additional peaks centred at 457.9 eV (2p$_{3/2}$) and 463.1 eV (2p$_{1/2}$) attributed to Ti$^{3+}$ ions. Previous research did not always report the presence of Ti$^{3+}$ after such treatments, implying a high stability of Ti$^{4+}$, which is against reduction to Ti$^{3+}$~\cite{Wang2011}. As shown in Figs.~\ref{fig:XPS}(a-c), after the hydrogenation at RT and 573 K, the Ti$^{4+}$(2p$_{3/2}$) peak moves to 459.5 eV and 458.7 eV, respectively in accordance with the previous results~\cite{Yan2014}. It has been reported that under certain circumstances Ti$^{2+}$ states can be observed in black titania~\cite{Panomsuwan2015,Singh2016}. By increasing temperature during the plasma treatment caused a Ti$^{2+}$ state to appear. This can be observed in Fig.~\ref{fig:XPS}(c), the new peaks emerged at 454.9 eV (2p$_{3/2}$) and 460.8 eV (2p$_{1/2}$) after hydrogen treatment at 573 K, which correspond to Ti$^{2+}$.

    Figs.~\ref{fig:XPS}(d-f) show the O 1s core level XPS spectra for the pristine and hydrogenated samples at RT and 573 K. The O 1s spectra of the pristine sample can be de-convoluted into two peaks at 530.2 eV and 532.1 eV. The first is ascribed to oxygen in Ti-O-Ti bonds and the second to oxygen in Ti-OH in accordance with earlier data~\cite{Fan2015,McCafferty1998}. There is evidence of changes in peak fractions of these two peaks representing Ti-O and Ti-OH bonds, although binding energies remain almost the same before and after hydrogenation treatments. However, a small positive shift in the binding energies of Ti-O-Ti after RT hydrogenation (from 530.1 to 530.9 eV) resembles the behaviour of Ti$^{4+}$(2p$_{3/2}$), where energies increased and returned after treatment at 573 K (from 458.9 eV to 459.5 eV). There is a third component on the surface, which has a binding energy of 533.5 eV and is assigned to O-C-O bond owing to unintentional carbon contamination during sample handling/plasma treatments.

    In addition, there is a clear trend of an increasing Ti-OH fraction and its further dominance over Ti-O-Ti after hydrogenation at 573 K. A similar tendency was observed and reported in other hydrogenated TiO$_2$~\cite{Chen2011}, which suggested the creation of more -OH bonds at subsurface regions. 

    Hydrogenated thin films show no significant disorder as presented in Fig.~\ref{fig:XRDUV}. Even though the same hydrogenation parameters on rutile samples were utilised for a longer time period a slightly amorphous phase of upper layers was detected by GIXRD. Based on the fact that there are no significant structural changes, and that the light absorption of the film after hydrogenation has been improved, we can deduce that the band gap narrowing took place with hydrogenation. Several suggestions on the origin of the band gap narrowing have been proposed. However, due to its complexity and mutual Ti interstitials and vacancies interplay, the reasons remain unclear~\cite{Chen2011,He2013}. Ti defects alone might play a small role, therefore one may guess that the localised-states at 0.7-1.0 eV and 0.92-1.37 eV are induced by oxygen vacancies and Ti-H bonds respectively~\cite{Naldoni2012,Zhou2013}. Researchers observed that the high electroconductivity in TiO$_2$ is due to Ti$^{3+}$ and Ti$^{2+}$ states along with oxygen vacancies and this is in accordance with XPS results, which show increasing of Ti defects with temperature of hydrogenation treatment~\cite{Amano2016,Eom2015}. However, such extreme changes in the electroconductivity have not been reported before. Typically, reduction from Ti$^{4+}$ to Ti$^{3+}$ originates from the surface and is always accompanied by a loss of oxygen.

    With increasing time/temperature of hydrogenation treatments, hydrogen tends to increase its number of bonds with TiO$_2$. This is demonstrated in the changes of the Ti-O-Ti/Ti-OH ratio with hydrogenation as shown in Figs.~\ref{fig:XPS}(a-c). 

    XPS results indicate the presence of both the Ti$^{2+}$ and Ti$^{3+}$ states, which are likely accompanied by V$_o$ defects in our TiO$_2$ films after hydrogenation treatment. We, therefore applied eMS to probe these defects and their temperature-dependent interactions in hydrogenated TiO$_2$ on the atomic scale.

 \subsection{Emission M{\"o}ssbauer Spectroscopy}

        \begin{figure}[ht]
                  \centering
                  \includegraphics[width=9cm, trim=0.5cm 2cm 2cm 1cm, clip]{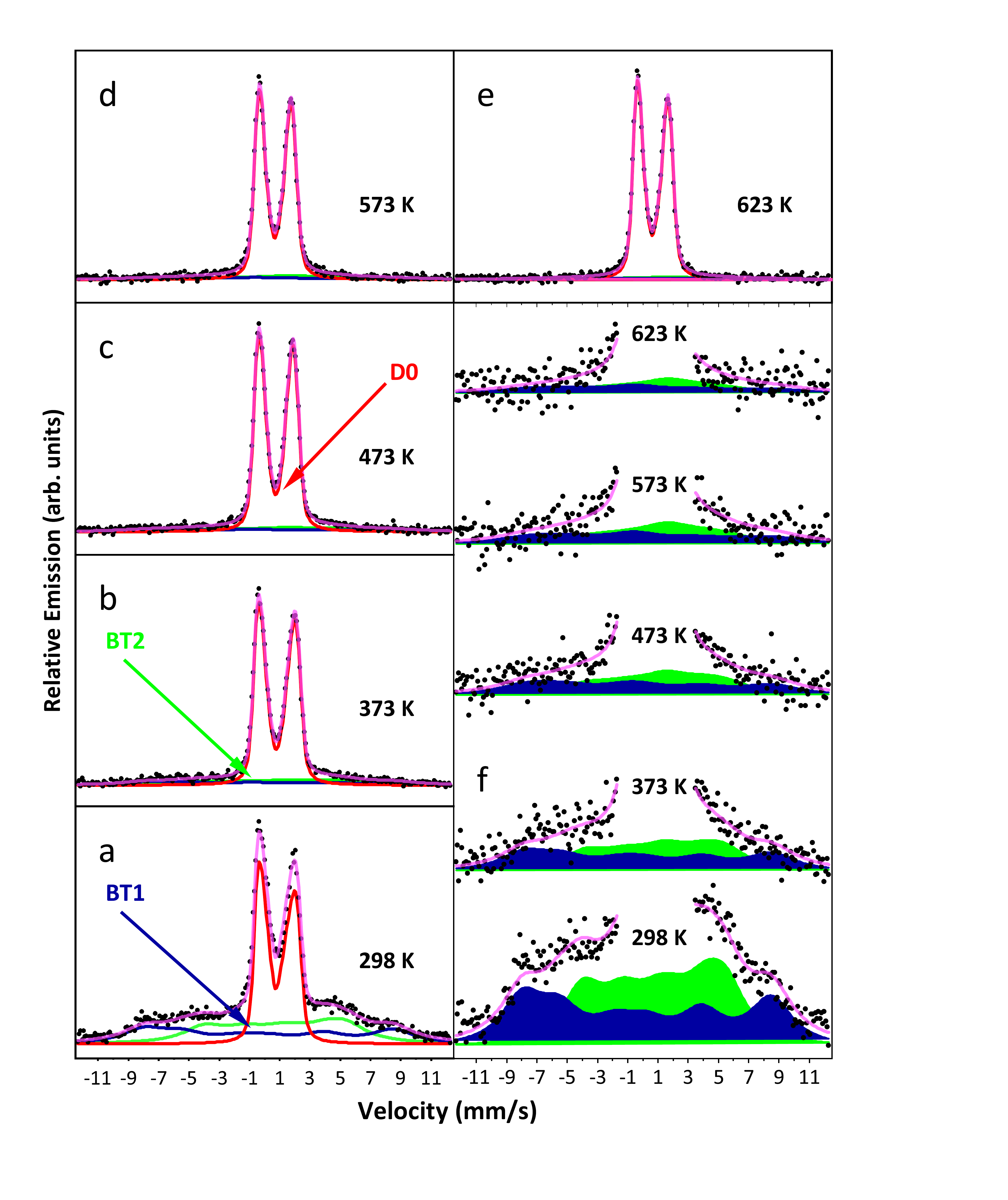}
                  \caption{(a-e) $^{57}$Fe emission M{\"o}ssbauer spectra obtained at the temperatures indicated after implantation of $^{57}$Mn into pristine anatase TiO$_2$. (f) Magnified view of the spectra showing the effect of the magnetic splitting.}
                  \label{fig:eMS1a}
                \end{figure}

        \begin{figure}[ht]
                  \centering
                  \includegraphics[width=9cm, trim=1.2cm 0.3cm 2cm 2cm, clip]{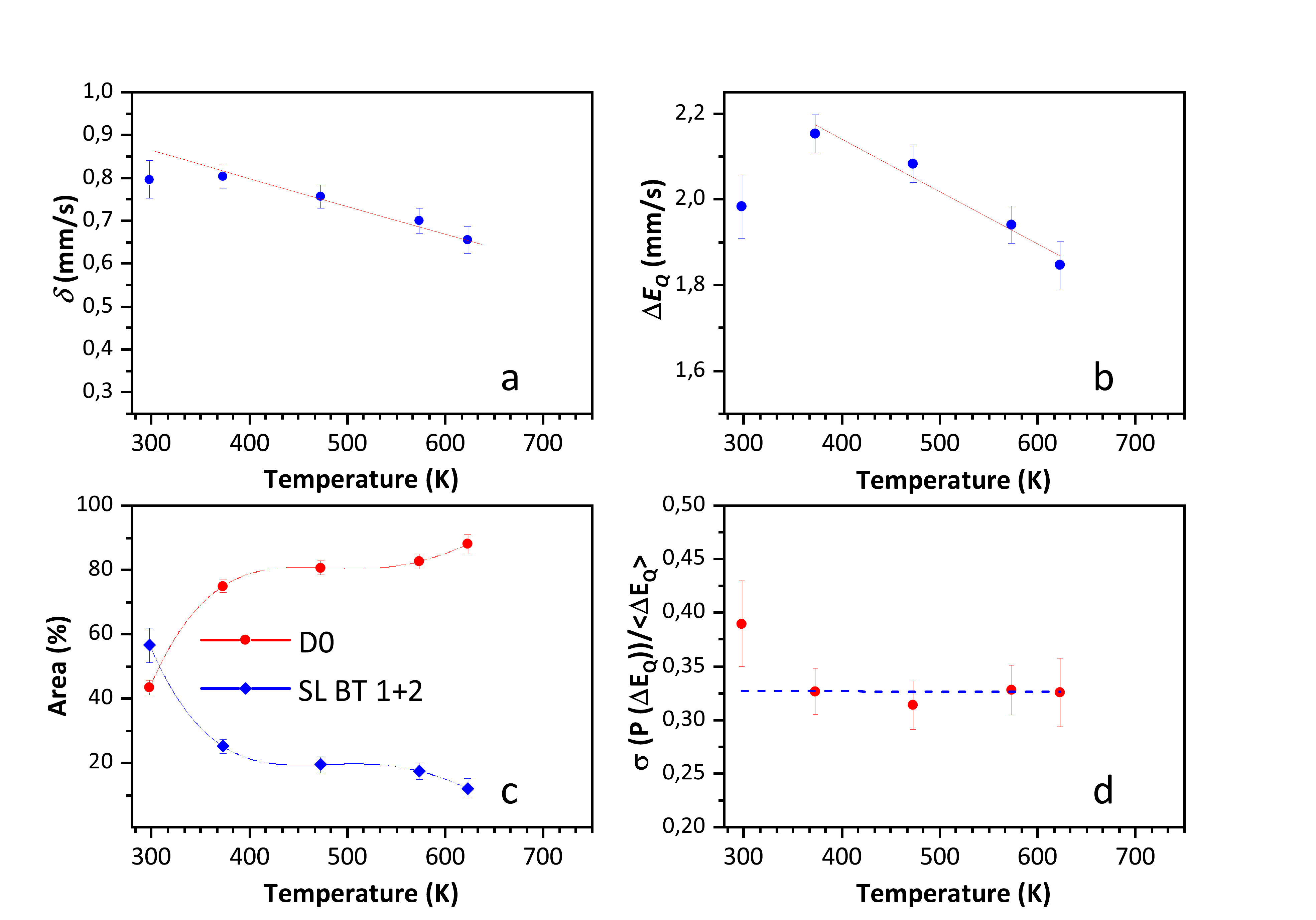}
                  \caption{Temperature dependence of hyperfine parameters of Fe$^{2+}$ (D0). (a) Average isomer shift. The solid line shows the second order Doppler shift. (b) Average quadrupole splitting. The solid line shows the linear trend above 373 K. (c) Area fractions of the spectral components of the spectra in Fig.~\ref{fig:eMS1a}. (d) Relative standard deviation of the quadrupole splitting distribution function. The dashed line shows the average above 373 K.}
                  \label{fig:eMS1b}
                \end{figure}

    A series of emission M{\"o}ssbauer spectra measured on the pristine sample is shown in Figs.~\ref{fig:eMS1a}(a-e), while Fig.~\ref{fig:eMS1a}(f) is focused on the magnetic features (wings). We evaluated the spectra in terms of three components by virtue of the \textsc{vinda} program~\cite{Gunnlaugsson2016}. The central part consists of a doublet due to high-spin Fe$^{2+}$ (D0). This part was calculated as a distribution of quadrupole splitting~\cite{Gunnlaugsson2006}. The quadrupole splitting and the isomer shift are coupled as $\delta$ = $\delta_0$ + $\delta_1 \cdot \Delta E_Q$, where $\delta_1$ and $\delta_0$ are fitting variables. The derived isomer shift and quadrupole splitting at RT are $\delta_{RT}$ = 0.8 $\pm$ 0.04 and $\Delta E_Q$ = 1.98 $\pm$ 0.07 mm/s, respectively. It is straight forward to assign D0 to Fe ions at the substitutional sites in anatase~\cite{Umek2014}.
    The wings of the spectra show magnetic hyperfine splitting, similar to those observed previously in the eMS studies on rutile TiO$_2$~\cite{Gunnlaugsson2014}, which were attributed to high-spin Fe$^{+3}$ showing spin-lattice relaxations on a comparable time-scale as the lifetime of the 14.4 keV M{\"o}ssbauer state of $^{57}$Fe. This part of the spectra was analyzed with the semi-empirical Blume-Tjon-based model (BT1 and 2)~\cite{Molholt2010}. Fig.~\ref{fig:eMS1b}(c) shows the relative area fractions of both Fe$^{2+}$ and Fe$^{3+}$ based on the simultaneous analysis of the spectra, while Figs.~\ref{fig:eMS1b}(a,b,d) concentrate on the hyperfine parameters of Fe$^{2+}$. The derived results reveal two annealing stages: 1) between 300 K and 373 K and 2) from 373 to 623 K. 

    During the first stage the majority of the Fe$^{3+}$ ions is transformed into Fe$^{2+}$ with increasing temperature and is clearly observable in Fig.~\ref{fig:eMS1b}(c). The quadrupole splitting value of Fe$^{2+}$ shown in Fig.~\ref{fig:eMS1b}(b) shifts, and the isomer shift demonstrates an increase relative to the second order Doppler shift (SOD) trend (in Fig.~\ref{fig:eMS1b}(a), shown as a solid line). The relative line broadening presented in Fig.~\ref{fig:eMS1b}(d) decreases, and the quadrupole splitting distribution becomes more symmetric, as observed through the coupling parameter $\delta_1$ between the isomer shift and the quadrupole splitting (not shown). This annealing stage is very similar to the annealing stage reported for rutile TiO$_2$~\cite{Gunnlaugsson2014} around 330 K. At this temperature the Fe$^{2+}$ area fraction, isomer shift and quadrupole splitting increased, while the quadrupole splitting distribution became narrower. We attribute this stage to the mobility of Ti interstitials (Ti$_{I}$) and/or Ti vacancies (V$_{Ti}$), leaving the probe atom in an environment with oxygen vacancies. Thus promoting the 2+ state over 3+ state. Gunnlaugsson \textit{et al.}~\cite{Gunnlaugsson2014} interpreted the annealing effect in terms of Ti$_{I}$, mostly because rutile tends to form these interstitials more easily than anatase~\cite{Morgan2010}. Due to this reason, the absence of the sharp annealing stage here at 330 K (unlike in rutile) can be in general interpreted as less noticeable contribution of V$_{Ti}$ and Ti$_{I}$ mutual annihilation~\cite{Morgan2010}. 

    There is no evidence of the $\sim$ 550-600 K annealing stage reported in rutile TiO$_2$~\cite{Gunnlaugsson2014}, which the authors attributed to the dissociation of Mn-V$_o$ pairs within the 1.45 min half-life of $^{57}$Mn. This suggests that Mn-V$_o$ pairs have higher dissociation energy in anatase, compared to rutile. Using the same method as in Ref.~\cite{Gunnlaugsson2014}, one finds a dissociation energy $>$ 1.8 eV for Mn-V$_o$ pairs in anatase TiO$_2$. Furthermore, the second annealing stage observed here with Fe$^{2+}$ (D0 component) dominating the spectra up to 650 K can be additionally supported by the fact that V$_o$ are easily formed in anatase, especially under ion radiation~\cite{Morgan2010,Pan2013}.

        \begin{figure}[ht]
                  \centering
                  \includegraphics[width=8cm, trim=2cm 2cm 2cm 2cm, clip]{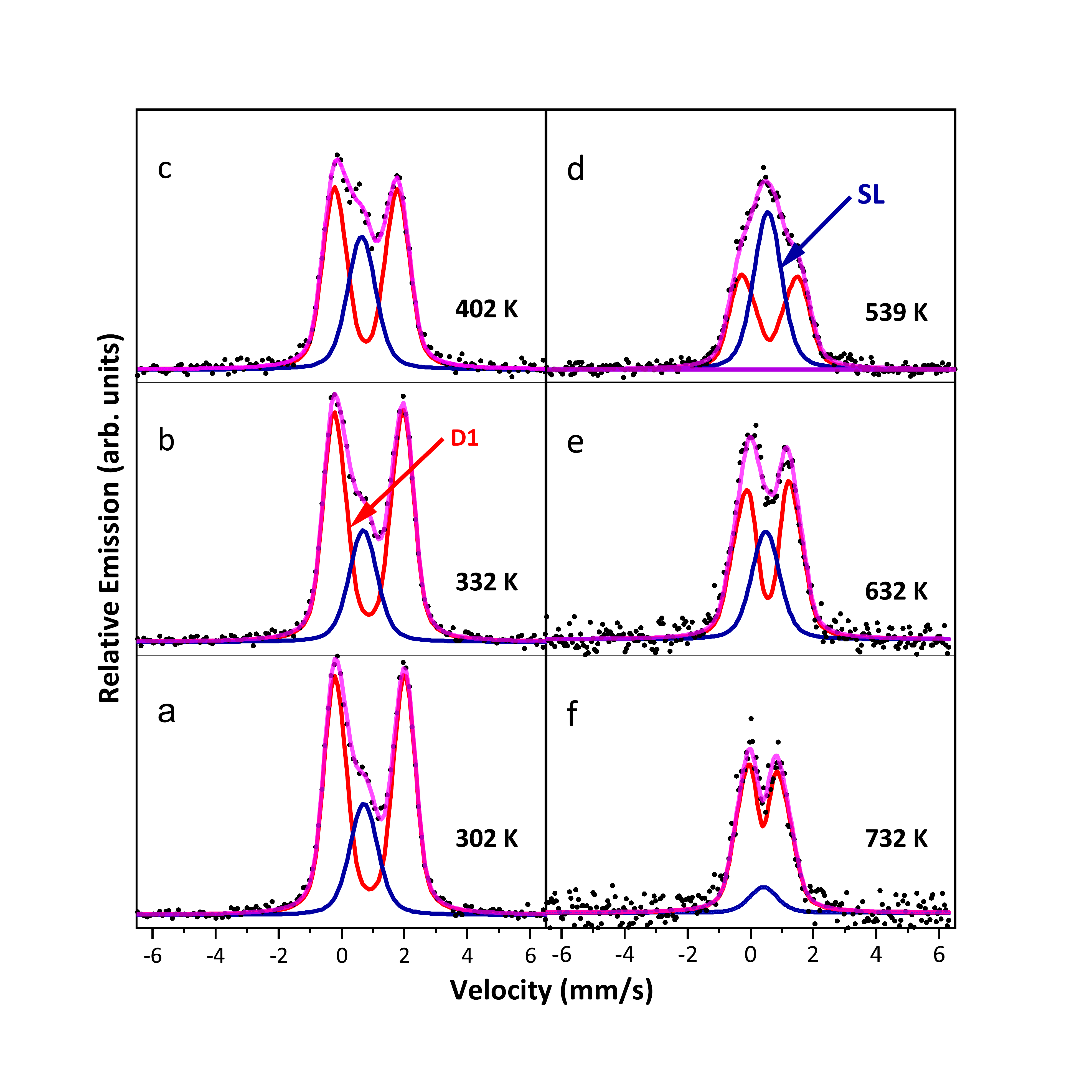}
                  \caption{$^{57}$Fe emission M{\"o}ssbauer spectra obtained at the temperatures indicated after implantation of $^{57}$Mn into TiO$_2$:H-RT.}
                  \label{fig:eMS2a}
                \end{figure}

    During the $^{57}$Mn $\rightarrow$ $^{57}$Fe $\beta^{-}$ decay, an average recoil energy of 40 eV is transferred to the $^{57}$Fe atom that may lead to a relocation of atoms into interstitial sites. Nonetheless, results show no indication of a spectral component that could be interpreted as a result of interstitial Ti (a threshold displacement energy of Ti atoms is around 39(1) eV~\cite{Robinson2014}). If interstitial Fe are formed, this could indicate that they can find their regular lattice site in less than 100 ns or they possess too low Debye-Waller factor to be observed in eMS measurements. The isomer shift shown in Fig.~\ref{fig:eMS1b}(a) follows roughly the SOD at T$>$ 370 K. The average quadrupole splitting in Fig.~\ref{fig:eMS1b}(b) shows a smooth dependency above 370 K, with a sudden increase from 300 to 370 K. This resembles the results observed in rutile, and is supposedly due to changes in the local environment~\cite{Gunnlaugsson2014}. Fig.~\ref{fig:eMS1b}(d) shows that the same trend with the width of the quadrupole splitting distribution, suggesting that Fe$^{2+}$ is not located at the amorphous environment in anatase. Only at higher doses of implantation (1$\times$10$^{16}$ ions/cm$^{2}$), does the anatase become amorphous~\cite{Ghicov2006}. 

        \begin{figure}[ht]
                  \centering
                  \includegraphics[width=9cm, trim=1.2cm 0.3cm 2cm 2cm, clip]{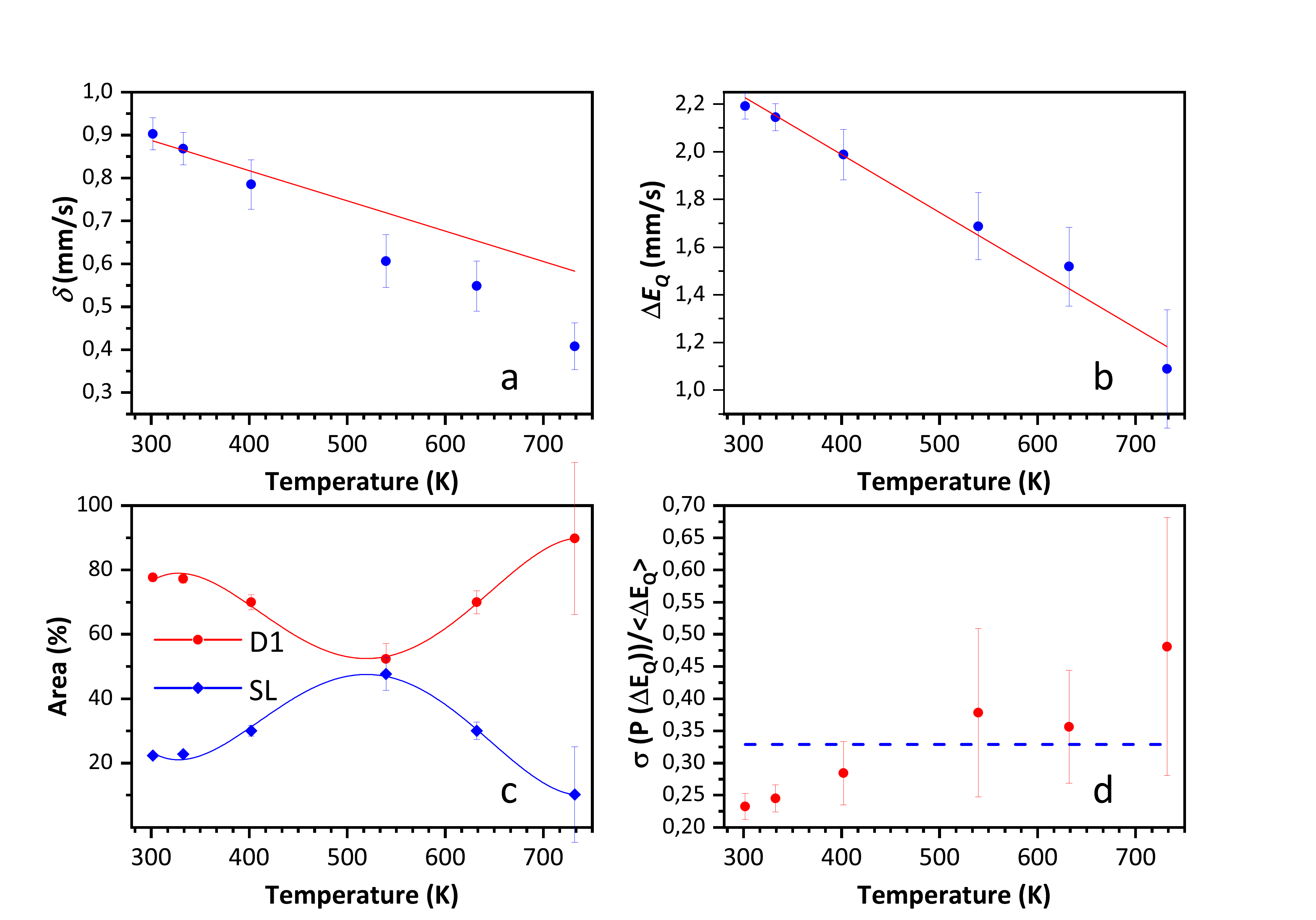}
                  \caption{Temperature dependence of hyperfine parameters of Fe$^{2+}$ (D1) for TiO$_2$:H-RT. (a) Average isomer shift. The solid line is a comparison to the SOD shift. (b) Average quadrupole splitting. (c) Area fractions of the spectral components of the spectra in Fig.~\ref{fig:eMS2a}. (d) Relative standard deviation of the quadrupole splitting distribution function.}
                  \label{fig:eMS2b}
                \end{figure}
    
    Fig.~\ref{fig:eMS2a} shows a series of eMS spectra obtained on the TiO$_2$:H-RT sample. There are two major differences in comparison with the spectra obtained on the pristine sample (see Fig.~\ref{fig:eMS1a}): 1) no Fe$^{3+}$ component is observed and 2) the spectra have additionally an apparent single line (SL) and a doublet (D1), which resembles D0 in pristine TiO$_2$ and assigned to a high spin Fe$^{2+}$. While the absence of Fe$^{3+}$ can be explained by the doping effects of hydrogen, the presence of a single line in the eMS spectra from non-cubic structure of anatase at first glance represents an unexpected feature. The hyperfine parameters and relative area fractions of Fe$^{2+}$ determined from the simultaneous analysis of the spectra are presented in Figs.~\ref{fig:eMS2b}(a-c). The derived isomer shift and quadrupole splitting for TiO$_2$:H-RT obtained at RT for D1 are $\delta_{RT}$ = 0.90 $\pm$ 0.04 and $\Delta E_Q$ = 2.19 $\pm$ 0.06 mm/s, respectively. Apparently, these hyperfine parameters resemble those obtained in rutile. 

    The unknown SL has an isomer shift of $\delta_{RT}$ = 0.72 $\pm$ 0.02 mm/s and demonstrates a temperature dependence. The isomer shift of D1 for the sample obtained after hydrogenation at RT depicted in Fig.~\ref{fig:eMS2b}(a) does not follow the SOD, and is characterised by $\delta_{RT}$ = 0.90$\pm$ 0.04 mm/s close to 300 K and by $\delta\sim 0.4$ mm/s close to 750 K. The $\delta_{RT}$ is similar to the previously obtained D0 in the pristine sample shown in Fig~\ref{fig:eMS1b}. This could either suggest a change in the local environment or addition of another unresolved spectral component with a lower isomer shift. The quadrupole splitting $\Delta E_Q$ of D1 is close to $\sim$ 2.2 mm/s, which is similar to the value obtained for D0 in the analysis of eMS spectra of the pristine sample above the 330 K annealing stage. For the pristine sample, the (extrapolated) quadrupole splitting of D0 is $\sim$ 1.8 mm/s at 700 K, while for D1 of the sample hydrogenated at RT, the value 1.2 mm/s is obtained. The presence of two unresolved quadrupole split components in Fig.~\ref{fig:eMS2b}(d) explains the broadening of the D1 component. At approximately 300 K, the spectra are dominated by a contribution similar to D0 of the pristine sample, while at 750 K it is also composed of another contribution that has lower isomer shift and quadrupole splitting. The addition of the two contributions would give a broader final component at elevated temperatures. This suggests a broad annealing stage centred around 500 K, coinciding with the peak in the SL area fraction as shown in Fig.~\ref{fig:eMS2b}(c). Moreover, there is no 330 K annealing stage observed in the pristine sample. 

     \begin{figure}[ht]
                  \centering
                  \includegraphics[width=8cm, trim=2cm 2cm 2cm 2cm, clip]{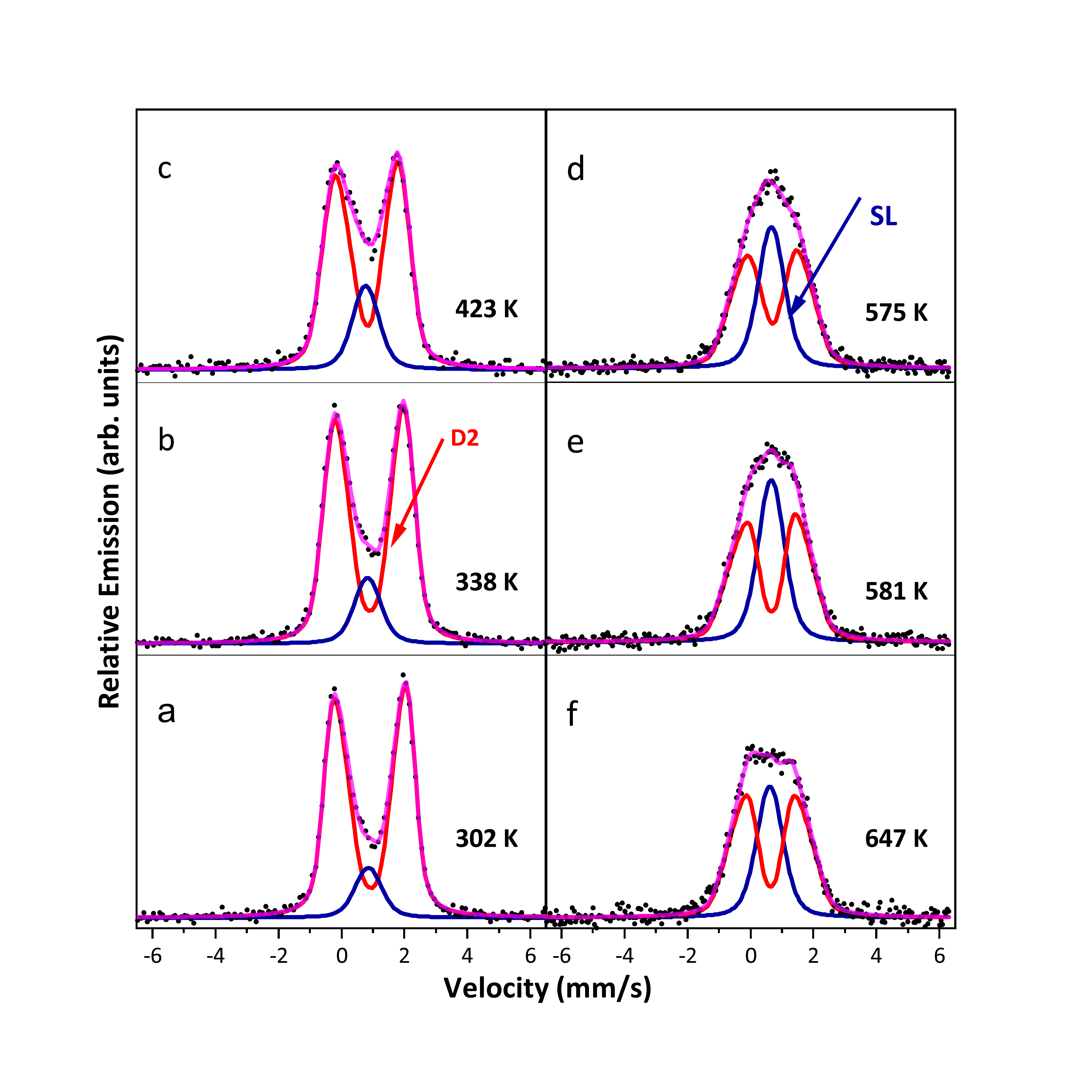}
                  \caption{$^{57}$Fe emission M{\"o}ssbauer spectra obtained at the temperatures indicated after implantation of $^{57}$Mn into TiO$_2$:H-573.}
                  \label{fig:eMS3a}
                \end{figure}

\begin{figure}[ht]
                  \centering
                  \includegraphics[width=9cm, trim=1.2cm 0.3cm 2cm 2cm, clip]{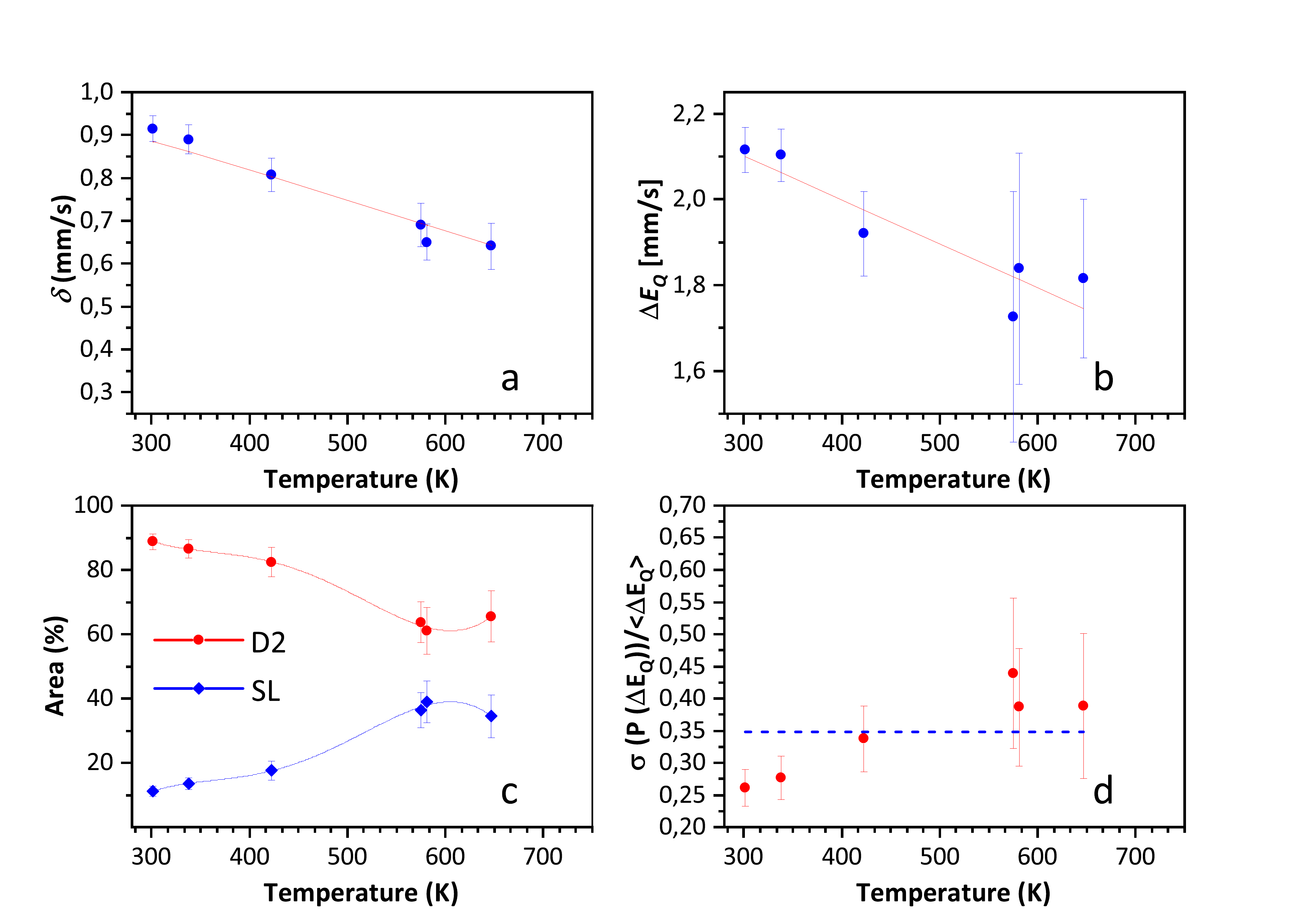}
                  \caption{Temperature dependence of hyperfine parameters of Fe$^{2+}$ (D2) for TiO$_2$:H-573. (a) Average isomer shift. The solid line is a fit to the SOD shift. (b) Average quadrupole splitting. (c) Area fractions of the spectral components of the spectra in Fig.~\ref{fig:eMS3a}. (d) Relative standard deviation of the quadrupole splitting distribution function.}
                  \label{fig:eMS3b}
                \end{figure}

    The sample hydrogenated at 573 K demonstrates a different temperature behaviour as depicted in Fig.~\ref{fig:eMS3a}. There is no annealing stage visible in the pristine sample. However, the collapsing stage persists up to 650 K as can be seen in Fig.~\ref{fig:eMS3b}(c). The isomer shift of D2 component shown in Fig.~\ref{fig:eMS3b}(a) has slightly higher values $\delta_{RT}$ = 0.92 $\pm$ 0.03 mm/s near 300 K and at 700 K (if extrapolated) reaches $\delta\sim$ 0.6 mm/s and seems to follow the SOD. The quadrupole splitting value is slightly lower at RT in comparison to TiO$_2$:H-RT. Although if it is extrapolated up to 700 K, it appears to be in order of 1.7 mm/s, which is significantly larger than in the previous sample (see Figs.~\ref{fig:eMS3b}(b) and ~\ref{fig:eMS2b}(b)). The influence of hydrogenation on the area fractions shows a steeper evolution of TiO$_2$:H-573 K features, where the SL contributes only 11.2\% of the spectral area (22.4\% in the case of TiO$_2$:H-RT) at RT. Here, the fractions (D1 and SL) almost reach a parity at 500 K. Afterwards, the SL decreases steadily which is different for the 573 K sample, where both components are stretched out for each other at 600 K. At this point we assume that the local atomic environment of D1 at RT is different to the one after the equivalent stage at 500 K (see Fig.~\ref{fig:eMS2b}). Although, Fe$^{2+}$ components seem to be similar in pristine and after hydrogenation near 300 K, the broadening of D2 component remains alike to TiO$_2$:H-RT sample as is obvious in Fig~\ref{fig:eMS3b}(d). 

    The hyperfine parameters change marginally after the hydrogenation treatment at 573 K. D2 component shows only a small growth of $\delta_{RT}$ = 0.92 $\pm$ 0.03 and decrease of $\Delta E_Q$ = 2.12 $\pm$ 0.05 mm/s. The increase of quadrupole splitting for the hydrogenated samples shows that the electric local environment around probes has become more asymmetrical. Additional $V_o$ and -OH bonds should increase the \textit{s}-electron density at the nucleus resulting in changes of isomer shift, which is reasonable since both act as shallow donors in TiO$_2$ even when H is isolated~\cite{Xiong2007,Li2014,Deak2011}. 

    The nature of the SL and absence of Fe$^{3+}$ requires more attention. To investigate this, one may need to study the local atomic environment further. At this point the results can be explained in terms of charge transfer. In the current system this transfer can be invoked by: 1) high amount of V$_o$ in the upper layers, that supports Fe$^{2+}$ state as well as the presence of Ti$^{2+}$ and Ti$^{3+}$ as detected by XPS measurements; 2) hydrogen bonds -OH; 3) both combined. Considering more than one degree of freedom, the annealing behavior of the treated samples could be slightly ambiguous. As V$_o$ undergo changes at low temperatures (at 360 K), we focus our understanding on point 2) hydrogen bonds - OH. On the grounds of a few experimental studies on metal oxides, the hydrogen impurities are expected to be in several forms~\cite{Dolci2007,ChenWan2013}. By and large, it is energetically favorable for hydrogen to occupy two states in bulk with different mobilities (i.e. activation energy). One form has a covalent bond to an oxygen atom (H$_{C}^{\bullet}$, placed perpendicular to $c$-axis), while the other behaves as a deep metastable interstitial (H$_{I}^{\bullet}$) or/and bound weakly via longer bonds to more than one neighboring oxygen atom~\cite{ChenWan2008}. According to previous studies, the charge state of H$_{I}^{\bullet}$ remains to be unknown, but often assumed to be positive~\cite{Kobayashi2012}. Results of another metal oxide WO$_3$ have shown that upon annealing hydrogen, which is weakly bound (or free), can leave as H$_2$, whereas -OH leaves in form of water~\cite{ChenWan2013}. The lower the concentration of hydrogen is, the higher is the energy needed (the jumping frequency lowers). Recent IR absorption study on anatase has revealed that the barrier energy for hydrogen motion is beyond 0.9 eV, which is two times higher than that for rutile~(0.53 eV)~\cite{Hupfer2017}. Thus, hydrogen in rutile cannot be accumulated in high quantities. Therefore it is likely stabilised by nearest defects. Bearing this in mind, it can be assumed that in the current case a possible synergistic configuration of higher amount of hydrogen and vacancies is responsible for the observed annealing behavior. 

    Due to the low activation energy, H$_{I}^{\bullet}$ can already start to slowly diffuse out at RT. If the latter does not develop that swiftly, a hopping of hydrogen among equivalent oxygen neighbors could occur with high rates. A proton can be located vibrating near an oxygen atom, then it jumps over the shortest bond lengths (smaller activation energy) to another oxygen and hops backwards to finalise the vibration near another oxygen~\cite{Sheikholeslam2016}. Our results show that implanted $^{57}$Fe ions replace Ti ions in the lattice, where the hydrogen hopping to various oxygen atoms happens to be in vicinity of the probe. Thus affecting the electric-field gradient direction rapidly, on the time-scale of the 14.4 keV lifetime of $^{57}$Fe M{\"o}ssbauer state. This can result in the occurrence of SL, which by all means is not a natural state in the tetragonal structure and can only be caused by dynamic processes. Similar tendency of H, for instance, was observed in SiO$_2$~\cite{Sheikholeslam2016}.
  
    With increasing temperature, more atoms participate in the motion, which could result in the formation of an identical defect occupying a different lattice position (i.e diffusion of the species). Nonetheless, if the activation energy is not sufficient for breaking up a chemical bond, it may lead to a transition (or reorientation) between equivalent positions in the lattice~\cite{Chaplygin2018}. For these reasons, one may need to consider that unbound-hydrogen has a significant impact on the electric field distribution around the probes.
    Studies performed by Lavrov \textit{et al.}~\cite{Lavrov2015} have shown that anatase (TiO$_2$) doped with hydrogen is unstable against annealing at temperatures above 570 K. Combining stepping annealing and XPS analysis alongside with nuclear reaction analysis studies have shown that in rutile~(TiO$_2$) hydrogen remains in the lattice up to 600 K, after which it depletes steeply~\cite{Nandasiri2015}. Therefore, one may conclude that hydrogen movement in anatase consists primarily of two steps, with the second step superimposing on the first at higher temperatures. Firstly, H$_{I}^{\bullet}$, which has a lower motion threshold, changes its orientation or moves throughout equivalent positions. This assertion can be justified by the eMS results in terms of the SL alteration with temperature. The eMS spectra obtained for the TiO$_2$:H-RT sample shows no temperature dependence up to 350-400 K, whereas from~$\approx$ 440 K a ``jumping''  of hydrogen throughout equivalent positions occurs. Secondly, when the activation energy is high enough the H$_{C}^{\bullet}$ covalent bonds dissociation is triggered. As the speed of desorption is always dependent on the concentration gradient, the recovery of area fractions begins when most of the hydrogen is depleted from upper layers, thus explaining the ``de-collapsing'' of SL after~$\approx$ 620 K. 

    The sample treated at 573 K exhibits a more perplexed scenario. Generally, it is similar to TiO$_2$:H-RT, however the collapsing stage is evident throughout the high temperature ranges. The phenomena could be explained in terms of ``hydrogen traps''. These traps are commonly formed when samples under study are with bulk defects, which in turn could be formed on a basis of V$_o$ agglomeration. Once these bulk defects are formed they prevent hydrogen atoms from swift diffusion and consequently from the early desorption~\cite{Nandasiri2015,Johnson1975}. 

    Besides, another scenario of two hydrogen interstitials forming a single H$_2$, can remain in the lattice under certain circumstances up to 973 K~\cite{Koch2014}. The gradual decrease in quadrupole splitting indicates a slow but steady recovery of defects. In this case, it is fair to expect the thermally-activated surface to bulk diffusion of Ti$^{3+}$ interstitials~\cite{Henderson1995}.
            
  \subsection{\textit{Ab-initio calculations}}

      To our knowledge, there have been solely few attempts to perform density functional theory (DFT) calculations on anatase doped with Fe, and only one study on changes in the electric field gradient (EFG) upon doping complemented with transmission M{\"o}ssbauer measurements. For the comparison with calculations, only samples doped with relatively high amount of $^{57}$Fe (2.8 and 5.4$\%$) were used, calculations performed with one Fe atom in the supercell could miss out possible lattice distortions~\cite{Rod2008}. This mismatch between theoretical approximations and experiments may cause a shift in the EFG that in fact could explain a typical 10-15$\%$ divergence of the experimental data from calculated results. Likewise, our experimental data does not match well with the early published results by Rodr\'{i}guez \textit{et al.}~\cite{Rod2008} (FLAPW with LSDA and GGA was used). Therefore it was necessary to perform our own calculations to check if a closer match to our experimental results can be obtained. 

      To simulate isolated impurities various approaches have been employed. Since our samples were implanted with dilute (ppm) amount of impurities, a $2\times2\times1$ supercell (SC) was utilised, taking into account that impurities did not interact with each other and structural relaxations were not affected. A tetragonal unit cell was featured with \textit{a = b =} 3.806 \text{\AA} and \textit{c =} 9.724 \text{\AA} constants taken from a 15 K neutron powder diffraction study~\cite{KORDATOS2018}. In the current situation it is important to predict and describe the charge state, structural, electronic, and hyperfine features of the impurity-host system. For that, calculations featuring various scenarios such as Fe being at a cationic site while surrounded with H, multiple configurations of equatorial/apical V$_o$ around the Fe impurity and with added electrons we performed. 

      The \textit{ab-initio} calculations were carried out employing the projector-augmented-waves (PAW) method implemented in the \textsc{vasp} environment. The theoretical Perdew–Burke-Ernzerhof (PBE) exchange-correlation functional with $3\times3\times3$ \textit{k}-points grids were used. Following our strategy a SC with \textit{a = b =} 7.612 \text{\AA} and \textit{c =} 9.724 \text{\AA} was constructed. For several cases the lattice parameters were optimised and only small differences were discovered in the results which are not presented here. In order to check the results we simultaneously calculated parameters for Fe:Ti and Fe:Ti(1$\bar{e}$) with an additional electron by means of \textsc{wien2k} (17.1), using the accurate (linearised) augmented plane wave plus the local orbitals ((L)APW+lo) method. The basis-set-size was defined by K$_{max}$ = $6/1.58$ = 3.8 a.u. $^{-1}$, and a $5\times5\times3$ \textit{k}-points mesh. Besides, two more possible sources of error were tested: the size of the supercell, and the use of the DFT+U approach to describe the \textit{d}-electrons of Fe and Ti. The quadrupole splitting (QS) values was calculated based on $\Delta E_Q = 6|A_Q|\sqrt{1+\eta^2/3}$ where the quadrupole coupling constant is given as follows: $A_Q = (ecQ V_{zz})/[4I_e(2I_e-1)E_0]$ with asymmetry parameter ${\eta=(V_{xx}-V_{yy})/V_{zz}}$. The nuclear electric quadrupole moment, $Q$, represents the first excited state of $^{57}$Fe, \textit{e} is the elementary charge and V$_{zz}$ is the EFG along the major axis (provides information on the symmetry of the electric field charge distribution)~\cite{ChenYa2007}. 

\begin{figure}[ht]
                    \centering
                    \includegraphics[width=7cm, trim=0cm 3cm 0cm 1cm, clip]{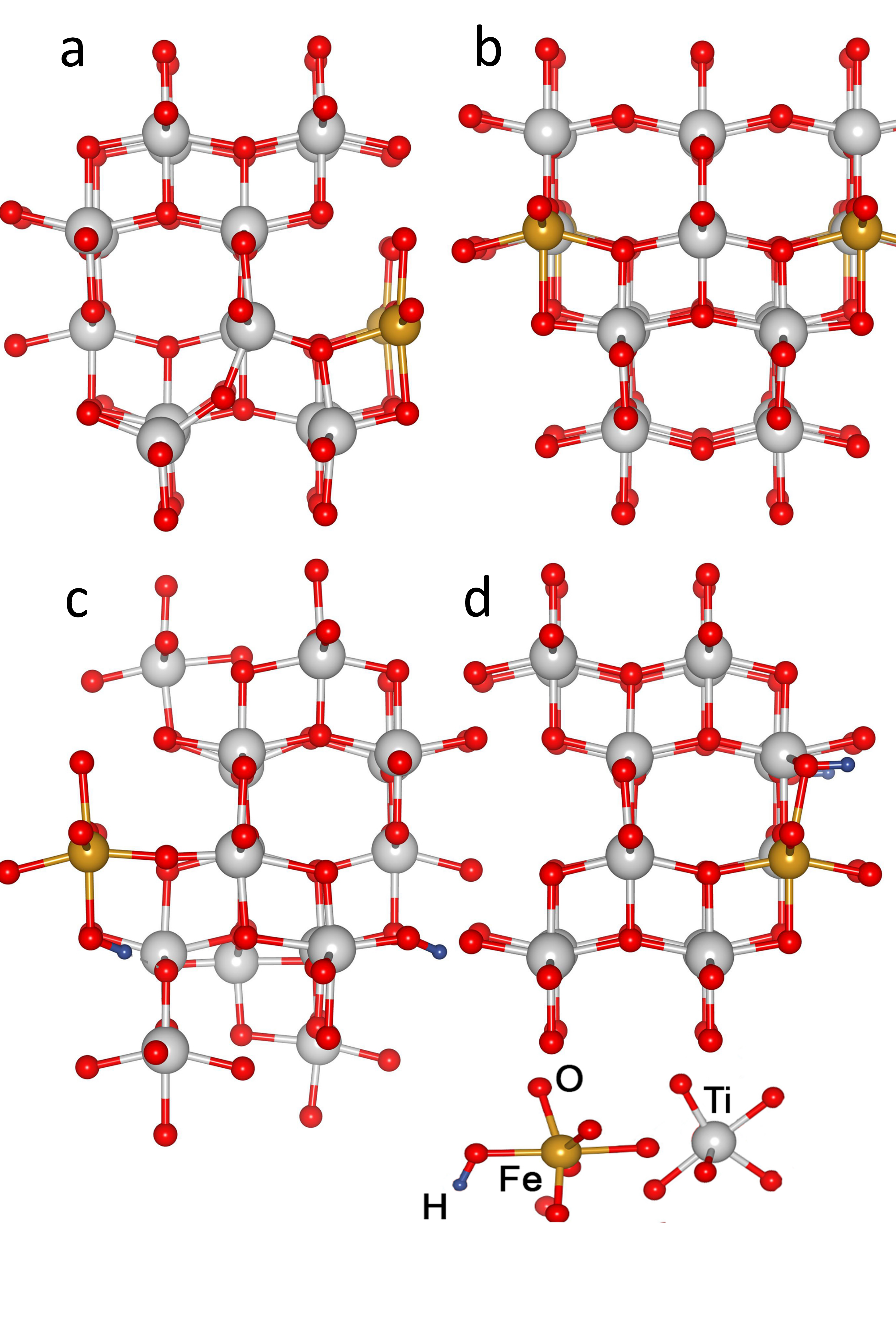}
                    \caption{Possible TiO$_2$ configurations reconstructed using \textit{ab-initio} calculations: (a) Fe:Ti+V$_o$ equatorial; (b) Fe:Ti+V$_o$ equatorial nnn.; (c) Fe:Ti+OH\#1; (d) Fe:Ti+OH\#2.}
                    \label{fig:DFTcells}
          \end{figure}

     Since eMS experiments were carried out with $^{57}$Mn/Fe it was straight forward to consider the Fe:Ti configuration as the starting point. For this reason, a case where a Ti atom was substituted by Fe, and additionally Fe:Ti(1$\bar{e}$) (checked with \textsc{vasp} \& \textsc{wien2k}) was reconstructed first. Secondly, Fe:Ti with either equatorial or apical V$_o$ (Figs.~\ref{fig:DFTcells}(a-b)) were reconstructed.  Last but not least, other calculations were implemented on the grounds of the Fe:Ti, additionally having hydrogen in two (OH\#1) and (OH\#2) configurations (Figs.~\ref{fig:DFTcells}c-d). It was assumed that OH\#1 could behave as H$_{I}^{\bullet}$, while OH\#2 could feature more robust bonding. The calculated parameters are listed in Table~\ref{tab:2}.

     Besides the aforementioned calculations of the EFG, an empirical model was applied to determine the M{\"o}ssbauer isomer shifts in ionic and covalent binary compounds. The model predicts room temperature isomer shift values for Fe$^{2+}$  in metal oxides. The empirical model relies mainly on the nearest-neighbour distances and Pauling electronegativity ($\Delta \chi_{P} <$ 2)~\cite{Gunnlaugsson2019}.

      \begin{table}[ht]
               \centering
                               \small
                  \caption{\label{tab:2}Calculations obtained for TiO$_2$ anatase. Here \textit{d}$_{AV}$ is the mean length of Fe-O bonds (\text{\AA}). V$_{zz}$  is the largest component  (given in 10$^{21}$ V$\cdot$m$^{-2}$) and $\eta$  is the asymmetry parameter of the EFG tensor. $\Delta E_Q$ and $\eta$ are given in mm/s. U3 or U5 eV reflect when a charge is added to \textit {d}-orbitals of Ti/Fe. Experimental data is for Fe$^{2+}$ component taken at RT. $\delta_{A}$ is the isomer shift (IS) based on the empirical model~\cite{Gunnlaugsson2019}. Mark~$^{\diamond}$ is added for a $3\times3\times1$ SC. Equatorial V$_o$ - eq.; nnn stands for a next-nearest-neighbour and m stands for a metastable state.}
            \begin{tabular}{c|c c c c c c >{\centering\arraybackslash}m{cm}}
                          \hline \hline
                               System:                                             &\textit{d}$_{AV}$             & V$_{zz}$      & $\eta$  & $\Delta E_Q$  & $\delta_{A}$\\ \hline

                               Fe:Ti$^{\diamond}$                                  & 1.936                        & 1.74                     & 0       & 0.309 &  0.954 \\
                               Fe:Ti+1$\bar{e}^{\diamond}$                         & 2.012                        & 2.17                     & 0       & 0.385 &  1.025 \\
                               Fe:Ti+OH\#1 m.                                      & 2.010                        & -5.50                    & 0.65    & 1.044 &  1.023 \\
                               Fe:Ti+OH\#2                                         & 2.031                        & -3.13                    & 0.97    & 0.638 &  1.043\\
                               Fe:Ti+V$_o$ eq.+U3                                  & 2.022                        & -7.25                    & 0.37    & 1.318 &  1.035\\
                               Fe:Ti+V$_o$ eq.+U5                                  & 2.026                        & -8.24                    & 0.45    & 1.514 &  1.038\\
                               Fe:Ti+V$_o$ nnn.                                    & 2.040                        & 9.69                     & 0.34    & 1.768 &  1.052\\
                               Fe:Ti+V$_o$ eq.nnn.$^{\diamond}$                    & 1.994                        & -14.7                    & 0.20    & 2.631 &  1.008\\ \hline
                               TiO$_2$ Pristine$_{exp}$ \textbf{D0}                &                        &                  &       & 1.98$\pm$0.01 & 0.80$\pm$0.04  \\
                               TiO$_2$:H-RT$_{exp}$     \textbf{D1}                &                        &                  &       & 2.19$\pm$0.06 & 0.90$\pm$0.04  \\
                               TiO$_2$:H-RT$_{exp}$     \textbf{SL}                &                        &                  &       & -             & 0.72$\pm$0.02  \\
                               TiO$_2$:H-573K$_{exp}$   \textbf{D2}                &                        &                  &       & 2.12$\pm$0.05 & 0.92$\pm$0.03  \\
                               TiO$_2$:H-573K$_{exp}$   \textbf{SL}                &                        &                  &       & -             & 0.86$\pm$0.02  \\
                              \hline
              \end{tabular}
          \end{table}

     The performed calculations show that after relaxation, local distortions affecting Fe:Ti bond lengths are anisotropic and higher than what have been observed with calculations performed by Rodr\'{i}guez-Torres \textit{et al.}~\cite{Rod2008}, however slightly lower (1.94~\text{\AA}) than the previously published XAFS data (Fe:Ti) (1.97~\text{\AA})~\cite{Sanyuan2007}. The difference in bond lengths could be related to the displacement which is symmetric in the PBE approximation. Furthermore, results for the calculated hyperfine parameters differ from ones obtained experimentally, similarly as in the previous study. On the basis of these calculations one may ponder that a combined configuration of Fe:Ti+OH\#2 with Fe:Ti+V$_o$ could result in close match with experimental data. In a charged supercell, the IS of a doublet likely remains within the range of Fe$^{3+}$. The trace of hydrogen breaks the symmetry of the EFG near the probe. Including V$_o$ alongside Fe doping in the calculation demonstrates that neither the high energetic stability nor results are close to the experimental data. Compared with the previous observation~\cite{Rod2008}, the current experimental results are in better agreement with our calculated results of Fe-O next-nearest-neighbor locations and QS match relatively close (1.77 with 1.98 mm/s, respectively) and the deviations are in order of 10-12$\%$. Moreover, the performed calculations are done at 0 K and if there is a low crystal field splitting near the probe then average values of the QS are used. Generally, for Fe$^{2+}$ there is always an additional electron in the system, which bears an extra part to the EFG and can jump between orbitals that likely cancels out the QS which is evident in eMS results shown in Figs.~\ref{fig:eMS2b} and \ref{fig:eMS3b}. Hence, V$_o$ induce Fe$^{2+}$ with reasonable QS values, while in the case of -OH configurations they are characterised with lower QS numbers, and likely are responsible for the SL. 

     At this end, one may assume that the high experimental values of QS obtained in the current work and in rutile~\cite{Gunnlaugsson2014} are due to the rich amount of V$_o$ and V$_{Ti}$ present in the samples, for neither the experimental results from the previous work nor the current work match precisely with the calculations. The empirical model applied for Fe$^{2+}$ seem to match closely the results obtained in the current study in comparison to the results of DFT calculations for IS (not presented here). However, a deviation in the order of $\pm$0.11 mm/s is observed for a pristine configuration Fe:Ti.

  \section{CONCLUSIONS}

     To summarise, hydrogenated anatase films show the presence of surface-bulk defects such as Ti-OH and V$_o$, which have a positive impact on light absorption and conductivity. $^{57}$Fe emission M{\"o}ssbauer spectroscopy following implantation of $^{57}$Mn ($t_{1/2}$ = 1.5 min) to a maximum local concentration of 10$^{-3}$ at.\% into pristine anatase films at various temperatures reveals two distinct annealing stages from RT up to 373 K and to 623 K. The first stage occurs due to a movement of vacancies and leave the probes in highly disturbed local atomic environment. The second stage is induced by mutual annihilation of V$_{Ti}$ and Ti$_I$. Pristine sample demonstrates a slow spin lattice relaxation of Fe$^{3+}$. The hydrogenated samples show a different behaviour depending on a degree of hydrogenation. In both cases there is no sign of Fe$^{3+}$ which is due to hydrogen being a shallow donor. Specifically, for the sample treated at RT, the interstitial hydrogen (with a small activation energy) starts hopping throughout equivalent positions around the probes even at RT. With increasing temperature, there is  a breakdown of the -OH bond and with further increasing temperature brings the additional amount of hydrogen into play. The latter manifests in eMS as the collapsing of a doublet into a broad single line. Nonetheless, for the sample treated at 573 K, these two annealing stages overlap each other due to the bulk hydrogen traps. This is manifested as the collapsed eMS spectra at high temperatures, because H$_2$ could be trapped in bulk defects up to elevated temperatures. 
     \textit{Ab-initio} studies suggest that the observed eMS results are most likely related to a vacancy configuration near the probe (Fe:Ti+V$_o$ eq.nnn.$^{\diamond}$). Regarding the hydrogenated anatase,
     additional \textit{ab-initio} studies are needed in order to reconstruct which position (hydrogen configuration) in the lattice could yield the EFG values close to the experimental results. Nevertheless, based on the current data one may presume that Fe:Ti+OH\#1 m. has the biggest influence on the EFG of SL. 

\begin{acknowledgments}
  Authors are indebted to the ISOLDE team, expecially to J. G. M. Correia for technical assistance during the beam time.
  This work was supported by the German Federal Ministry of Education and Research (BMBF) projects 05K16SI1, 05K19SI1 and 05K16PGA. Additional funding is from the European Union's Horizon 2020 research and innovation programme under grant agreement no. 654002 (ENSAR2). The authors are also grateful to CERN/ISOLDE for support of the experiment IS653. B. Qi and S. {\'O}lafsson acknowledge the support from the Icelandic University Research Fund. I. Unzueta acknowledges the support of Ministry of Economy and Competitiveness (MINECO/FEDER) under the project RTI2018-094683-B-C55 and Basque Government Grant IT-1005-16. D. Naidoo, K. Bharuth-Ram, H. Masenda and G. Peters acknowledge support of the DST/NRF, South Africa. We are thankful for research funding by the state of Thuringia and the European Union within the frame of the European Funds for Regional Development (EFRD) under grant 12021-715.
\end{acknowledgments}

\bibliography{biblio}

\begin{thebibliography}{71}%
\makeatletter
\providecommand \@ifxundefined [1]{%
 \@ifx{#1\undefined}
}%
\providecommand \@ifnum [1]{%
 \ifnum #1\expandafter \@firstoftwo
 \else \expandafter \@secondoftwo
 \fi
}%
\providecommand \@ifx [1]{%
 \ifx #1\expandafter \@firstoftwo
 \else \expandafter \@secondoftwo
 \fi
}%
\providecommand \natexlab [1]{#1}%
\providecommand \enquote  [1]{``#1''}%
\providecommand \bibnamefont  [1]{#1}%
\providecommand \bibfnamefont [1]{#1}%
\providecommand \citenamefont [1]{#1}%
\providecommand \href@noop [0]{\@secondoftwo}%
\providecommand \href [0]{\begingroup \@sanitize@url \@href}%
\providecommand \@href[1]{\@@startlink{#1}\@@href}%
\providecommand \@@href[1]{\endgroup#1\@@endlink}%
\providecommand \@sanitize@url [0]{\catcode `\\12\catcode `\$12\catcode
  `\&12\catcode `\#12\catcode `\^12\catcode `\_12\catcode `\%12\relax}%
\providecommand \@@startlink[1]{}%
\providecommand \@@endlink[0]{}%
\providecommand \url  [0]{\begingroup\@sanitize@url \@url }%
\providecommand \@url [1]{\endgroup\@href {#1}{\urlprefix }}%
\providecommand \urlprefix  [0]{URL }%
\providecommand \Eprint [0]{\href }%
\providecommand \doibase [0]{https://doi.org/}%
\providecommand \selectlanguage [0]{\@gobble}%
\providecommand \bibinfo  [0]{\@secondoftwo}%
\providecommand \bibfield  [0]{\@secondoftwo}%
\providecommand \translation [1]{[#1]}%
\providecommand \BibitemOpen [0]{}%
\providecommand \bibitemStop [0]{}%
\providecommand \bibitemNoStop [0]{.\EOS\space}%
\providecommand \EOS [0]{\spacefactor3000\relax}%
\providecommand \BibitemShut  [1]{\csname bibitem#1\endcsname}%
\let\auto@bib@innerbib\@empty
\bibitem [{\citenamefont {Stevanovic}\ and\ \citenamefont
  {Yates}(2013)}]{Stevanovic2013}%
  \BibitemOpen
  \bibfield  {author} {\bibinfo {author} {\bibfnamefont {A.}~\bibnamefont
  {Stevanovic}}\ and\ \bibinfo {author} {\bibfnamefont {J.~T.}\ \bibnamefont
  {Yates}},\ }\bibfield  {title} {\bibinfo {title} {Electron {H}opping through
  {T}i{O}$_2$ {P}owder: {A} {S}tudy by {P}hotoluminescence {S}pectroscopy},\
  }\href {https://doi.org/10.1021/jp407765r} {\bibfield  {journal} {\bibinfo
  {journal} {J. Phys. Chem. C}\ }\textbf {\bibinfo {volume} {117}},\ \bibinfo
  {pages} {24189} (\bibinfo {year} {2013})}\BibitemShut {NoStop}%
\bibitem [{\citenamefont {Stevanovic}\ \emph {et~al.}(2012)\citenamefont
  {Stevanovic}, \citenamefont {B{\"u}ttner}, \citenamefont {Zhang},\ and\
  \citenamefont {Yates}}]{Stevanovic2012}%
  \BibitemOpen
  \bibfield  {author} {\bibinfo {author} {\bibfnamefont {A.}~\bibnamefont
  {Stevanovic}}, \bibinfo {author} {\bibfnamefont {M.}~\bibnamefont
  {B{\"u}ttner}}, \bibinfo {author} {\bibfnamefont {Z.}~\bibnamefont {Zhang}},\
  and\ \bibinfo {author} {\bibfnamefont {J.~T.}\ \bibnamefont {Yates}},\
  }\bibfield  {title} {\bibinfo {title} {Photoluminescence of {T}i{O}$_2$:
  {E}ffect of {UV} {L}ight and {A}dsorbed {M}olecules on {S}urface {B}and
  {S}tructure},\ }\href {https://doi.org/10.1021/ja2072737} {\bibfield
  {journal} {\bibinfo  {journal} {J. Am. Chem. Soc.}\ }\textbf {\bibinfo
  {volume} {134}},\ \bibinfo {pages} {324} (\bibinfo {year}
  {2012})}\BibitemShut {NoStop}%
\bibitem [{\citenamefont {Kurian}\ \emph {et~al.}(2013)\citenamefont {Kurian},
  \citenamefont {Seo},\ and\ \citenamefont {Jeon}}]{Kurian2013}%
  \BibitemOpen
  \bibfield  {author} {\bibinfo {author} {\bibfnamefont {S.}~\bibnamefont
  {Kurian}}, \bibinfo {author} {\bibfnamefont {H.}~\bibnamefont {Seo}},\ and\
  \bibinfo {author} {\bibfnamefont {H.}~\bibnamefont {Jeon}},\ }\bibfield
  {title} {\bibinfo {title} {Significant {E}nhancement in {V}isible {L}ight
  {A}bsorption of {T}i{O}$_2$ {N}anotube {A}rrays by {S}urface {B}and {G}ap
  {T}uning},\ }\href {https://doi.org/10.1021/jp405207e} {\bibfield  {journal}
  {\bibinfo  {journal} {J. Phys. Chem. C}\ }\textbf {\bibinfo {volume} {117}},\
  \bibinfo {pages} {16811} (\bibinfo {year} {2013})}\BibitemShut {NoStop}%
\bibitem [{\citenamefont {Lu}\ \emph {et~al.}(2011)\citenamefont {Lu},
  \citenamefont {Dai}, \citenamefont {Jin},\ and\ \citenamefont
  {Huang}}]{Lu2011}%
  \BibitemOpen
  \bibfield  {author} {\bibinfo {author} {\bibfnamefont {J.}~\bibnamefont
  {Lu}}, \bibinfo {author} {\bibfnamefont {Y.}~\bibnamefont {Dai}}, \bibinfo
  {author} {\bibfnamefont {H.}~\bibnamefont {Jin}},\ and\ \bibinfo {author}
  {\bibfnamefont {B.}~\bibnamefont {Huang}},\ }\bibfield  {title} {\bibinfo
  {title} {Effective increasing of optical absorption and energy conversion
  efficiency of anatase {T}i{O}$_2$ nanocrystals by hydrogenation},\ }\href
  {https://doi.org/10.1039/C1CP22726B} {\bibfield  {journal} {\bibinfo
  {journal} {Phys. Chem. Chem. Phys.}\ }\textbf {\bibinfo {volume} {13}},\
  \bibinfo {pages} {18063} (\bibinfo {year} {2011})}\BibitemShut {NoStop}%
\bibitem [{\citenamefont {Hoffmann}\ \emph {et~al.}(1995)\citenamefont
  {Hoffmann}, \citenamefont {Martin}, \citenamefont {Choi},\ and\ \citenamefont
  {Bahnemann}}]{Hoffmann1995}%
  \BibitemOpen
  \bibfield  {author} {\bibinfo {author} {\bibfnamefont {M.~R.}\ \bibnamefont
  {Hoffmann}}, \bibinfo {author} {\bibfnamefont {S.~T.}\ \bibnamefont
  {Martin}}, \bibinfo {author} {\bibfnamefont {W.}~\bibnamefont {Choi}},\ and\
  \bibinfo {author} {\bibfnamefont {D.~W.}\ \bibnamefont {Bahnemann}},\
  }\bibfield  {title} {\bibinfo {title} {Environmental {A}pplications of
  {S}emiconductor {P}hotocatalysis},\ }\href
  {https://doi.org/10.1021/cr00033a004} {\bibfield  {journal} {\bibinfo
  {journal} {Chem. Rev.}\ }\textbf {\bibinfo {volume} {95}},\ \bibinfo {pages}
  {69} (\bibinfo {year} {1995})}\BibitemShut {NoStop}%
\bibitem [{\citenamefont {Chen}\ \emph {et~al.}(2011)\citenamefont {Chen},
  \citenamefont {Liu}, \citenamefont {Yu},\ and\ \citenamefont
  {Mao}}]{Chen2011}%
  \BibitemOpen
  \bibfield  {author} {\bibinfo {author} {\bibfnamefont {X.}~\bibnamefont
  {Chen}}, \bibinfo {author} {\bibfnamefont {L.}~\bibnamefont {Liu}}, \bibinfo
  {author} {\bibfnamefont {P.~Y.}\ \bibnamefont {Yu}},\ and\ \bibinfo {author}
  {\bibfnamefont {S.~S.}\ \bibnamefont {Mao}},\ }\bibfield  {title} {\bibinfo
  {title} {Increasing {S}olar {A}bsorption for {P}hotocatalysis with {B}lack
  hydrogenated {T}itanium {D}ioxide {N}anocrystals},\ }\href
  {https://doi.org/10.1126/science.1200448} {\bibfield  {journal} {\bibinfo
  {journal} {Science}\ }\textbf {\bibinfo {volume} {331}},\ \bibinfo {pages}
  {746} (\bibinfo {year} {2011})}\BibitemShut {NoStop}%
\bibitem [{\citenamefont {Xia}\ and\ \citenamefont {Chen}(2013)}]{Xia2013}%
  \BibitemOpen
  \bibfield  {author} {\bibinfo {author} {\bibfnamefont {T.}~\bibnamefont
  {Xia}}\ and\ \bibinfo {author} {\bibfnamefont {X.}~\bibnamefont {Chen}},\
  }\bibfield  {title} {\bibinfo {title} {Revealing the structural properties of
  hydrogenated black {T}i{O}$_2$ nanocrystals},\ }\href
  {https://doi.org/10.1039/C3TA01589K} {\bibfield  {journal} {\bibinfo
  {journal} {J. Mater. Chem. A}\ }\textbf {\bibinfo {volume} {1}},\ \bibinfo
  {pages} {2983} (\bibinfo {year} {2013})}\BibitemShut {NoStop}%
\bibitem [{\citenamefont {Wang}\ \emph
  {et~al.}(2019{\natexlab{a}})\citenamefont {Wang}, \citenamefont {Xiong},
  \citenamefont {Cheng}, \citenamefont {Chen}, \citenamefont {Kups},
  \citenamefont {Wang},\ and\ \citenamefont {Schaaf}}]{HWang2019}%
  \BibitemOpen
  \bibfield  {author} {\bibinfo {author} {\bibfnamefont {H.}~\bibnamefont
  {Wang}}, \bibinfo {author} {\bibfnamefont {J.}~\bibnamefont {Xiong}},
  \bibinfo {author} {\bibfnamefont {X.}~\bibnamefont {Cheng}}, \bibinfo
  {author} {\bibfnamefont {G.}~\bibnamefont {Chen}}, \bibinfo {author}
  {\bibfnamefont {T.}~\bibnamefont {Kups}}, \bibinfo {author} {\bibfnamefont
  {D.}~\bibnamefont {Wang}},\ and\ \bibinfo {author} {\bibfnamefont
  {P.}~\bibnamefont {Schaaf}},\ }\bibfield  {title} {\bibinfo {title} {N-doped
  {T}i{O}$_2$ with a disordered surface layer fabricated via plasma treatment
  as an anode with clearly enhanced performance for rechargeable sodium ion
  batteries},\ }\href {https://doi.org/10.1039/C9SE00350A} {\bibfield
  {journal} {\bibinfo  {journal} {Sustainable Energy Fuels}\ ,\ } (\bibinfo
  {year} {2019}{\natexlab{a}})}\BibitemShut {NoStop}%
\bibitem [{\citenamefont {Wang}\ \emph
  {et~al.}(2019{\natexlab{b}})\citenamefont {Wang}, \citenamefont {Mayrhofer},
  \citenamefont {Hoefer}, \citenamefont {Estrade}, \citenamefont
  {Lopez-Conesa}, \citenamefont {Zhou}, \citenamefont {Lin}, \citenamefont
  {Peiró}, \citenamefont {Fan}, \citenamefont {Shen}, \citenamefont
  {Schaefer}, \citenamefont {Moseler}, \citenamefont {Braeuer},\ and\
  \citenamefont {Waag}}]{Wang2019}%
  \BibitemOpen
  \bibfield  {author} {\bibinfo {author} {\bibfnamefont {X.}~\bibnamefont
  {Wang}}, \bibinfo {author} {\bibfnamefont {L.}~\bibnamefont {Mayrhofer}},
  \bibinfo {author} {\bibfnamefont {M.}~\bibnamefont {Hoefer}}, \bibinfo
  {author} {\bibfnamefont {S.}~\bibnamefont {Estrade}}, \bibinfo {author}
  {\bibfnamefont {L.}~\bibnamefont {Lopez-Conesa}}, \bibinfo {author}
  {\bibfnamefont {H.}~\bibnamefont {Zhou}}, \bibinfo {author} {\bibfnamefont
  {Y.}~\bibnamefont {Lin}}, \bibinfo {author} {\bibfnamefont {F.}~\bibnamefont
  {Peiró}}, \bibinfo {author} {\bibfnamefont {Z.}~\bibnamefont {Fan}},
  \bibinfo {author} {\bibfnamefont {H.}~\bibnamefont {Shen}}, \bibinfo {author}
  {\bibfnamefont {L.}~\bibnamefont {Schaefer}}, \bibinfo {author}
  {\bibfnamefont {M.}~\bibnamefont {Moseler}}, \bibinfo {author} {\bibfnamefont
  {G.}~\bibnamefont {Braeuer}},\ and\ \bibinfo {author} {\bibfnamefont
  {A.}~\bibnamefont {Waag}},\ }\bibfield  {title} {\bibinfo {title} {Facile and
  efficient atomic hydrogenation enabled black tio2 with enhanced
  photo-electrochemical activity via a favorably low-energy-barrier pathway},\
  }\href {https://doi.org/10.1002/aenm.201900725} {\bibfield  {journal}
  {\bibinfo  {journal} {Adv. Energy Mater.}\ }\textbf {\bibinfo {volume} {9}},\
  \bibinfo {pages} {1900725} (\bibinfo {year}
  {2019}{\natexlab{b}})}\BibitemShut {NoStop}%
\bibitem [{\citenamefont {Syzgantseva}\ \emph {et~al.}(2011)\citenamefont
  {Syzgantseva}, \citenamefont {Gonzalez-Navarrete}, \citenamefont {Calatayud},
  \citenamefont {Bromley},\ and\ \citenamefont {Minot}}]{Syzgantseva2011}%
  \BibitemOpen
  \bibfield  {author} {\bibinfo {author} {\bibfnamefont {O.~A.}\ \bibnamefont
  {Syzgantseva}}, \bibinfo {author} {\bibfnamefont {P.}~\bibnamefont
  {Gonzalez-Navarrete}}, \bibinfo {author} {\bibfnamefont {M.}~\bibnamefont
  {Calatayud}}, \bibinfo {author} {\bibfnamefont {S.}~\bibnamefont {Bromley}},\
  and\ \bibinfo {author} {\bibfnamefont {C.}~\bibnamefont {Minot}},\ }\bibfield
   {title} {\bibinfo {title} {Theoretical investigation of the hydrogenation of
  (tio2)n clusters (n = 1–10)},\ }\href {https://doi.org/10.1021/jp2050349}
  {\bibfield  {journal} {\bibinfo  {journal} {J. Phys. Chem. C}\ }\textbf
  {\bibinfo {volume} {115}},\ \bibinfo {pages} {15890} (\bibinfo {year}
  {2011})}\BibitemShut {NoStop}%
\bibitem [{\citenamefont {Aschauer}\ and\ \citenamefont
  {Selloni}(2012)}]{Aschauer2012}%
  \BibitemOpen
  \bibfield  {author} {\bibinfo {author} {\bibfnamefont {U.}~\bibnamefont
  {Aschauer}}\ and\ \bibinfo {author} {\bibfnamefont {A.}~\bibnamefont
  {Selloni}},\ }\bibfield  {title} {\bibinfo {title} {Hydrogen interaction with
  the anatase tio2(101) surface},\ }\href {https://doi.org/10.1039/C2CP42288C}
  {\bibfield  {journal} {\bibinfo  {journal} {Phys. Chem. Chem. Phys.}\
  }\textbf {\bibinfo {volume} {14}},\ \bibinfo {pages} {16595} (\bibinfo {year}
  {2012})}\BibitemShut {NoStop}%
\bibitem [{\citenamefont {Chen}\ \emph {et~al.}(2015)\citenamefont {Chen},
  \citenamefont {Liu},\ and\ \citenamefont {Huang}}]{Chen2015}%
  \BibitemOpen
  \bibfield  {author} {\bibinfo {author} {\bibfnamefont {X.}~\bibnamefont
  {Chen}}, \bibinfo {author} {\bibfnamefont {L.}~\bibnamefont {Liu}},\ and\
  \bibinfo {author} {\bibfnamefont {F.}~\bibnamefont {Huang}},\ }\bibfield
  {title} {\bibinfo {title} {Black titanium dioxide ({T}i{O}$_2$)
  nanomaterials},\ }\href {https://doi.org/10.1039/C4CS00330F} {\bibfield
  {journal} {\bibinfo  {journal} {Chem. Soc. Rev.}\ }\textbf {\bibinfo {volume}
  {44}},\ \bibinfo {pages} {1861} (\bibinfo {year} {2015})}\BibitemShut
  {NoStop}%
\bibitem [{\citenamefont {Xiong}\ \emph {et~al.}(2007)\citenamefont {Xiong},
  \citenamefont {Robertson},\ and\ \citenamefont {Clark}}]{Xiong2007}%
  \BibitemOpen
  \bibfield  {author} {\bibinfo {author} {\bibfnamefont {K.}~\bibnamefont
  {Xiong}}, \bibinfo {author} {\bibfnamefont {J.}~\bibnamefont {Robertson}},\
  and\ \bibinfo {author} {\bibfnamefont {S.~J.}\ \bibnamefont {Clark}},\
  }\bibfield  {title} {\bibinfo {title} {Behavior of hydrogen in wide band gap
  oxides},\ }\href {https://doi.org/10.1063/1.2798910} {\bibfield  {journal}
  {\bibinfo  {journal} {J. Appl. Phys.}\ }\textbf {\bibinfo {volume} {102}},\
  \bibinfo {pages} {083710} (\bibinfo {year} {2007})}\BibitemShut {NoStop}%
\bibitem [{\citenamefont {Li}\ and\ \citenamefont {Robertson}(2014)}]{Li2014}%
  \BibitemOpen
  \bibfield  {author} {\bibinfo {author} {\bibfnamefont {H.}~\bibnamefont
  {Li}}\ and\ \bibinfo {author} {\bibfnamefont {J.}~\bibnamefont {Robertson}},\
  }\bibfield  {title} {\bibinfo {title} {Behaviour of hydrogen in wide band gap
  oxides},\ }\href {https://doi.org/10.1063/1.4878415} {\bibfield  {journal}
  {\bibinfo  {journal} {J. Appl. Phys.}\ }\textbf {\bibinfo {volume} {115}},\
  \bibinfo {pages} {203708} (\bibinfo {year} {2014})}\BibitemShut {NoStop}%
\bibitem [{\citenamefont {Brant}\ \emph {et~al.}(2011)\citenamefont {Brant},
  \citenamefont {Yang}, \citenamefont {Giles},\ and\ \citenamefont
  {Halliburton}}]{Brant2011}%
  \BibitemOpen
  \bibfield  {author} {\bibinfo {author} {\bibfnamefont {A.~T.}\ \bibnamefont
  {Brant}}, \bibinfo {author} {\bibfnamefont {S.}~\bibnamefont {Yang}},
  \bibinfo {author} {\bibfnamefont {N.~C.}\ \bibnamefont {Giles}},\ and\
  \bibinfo {author} {\bibfnamefont {L.~E.}\ \bibnamefont {Halliburton}},\
  }\bibfield  {title} {\bibinfo {title} {Hydrogen donors and ti$^{3+}$ ions in
  reduced tio$_2$ crystals},\ }\href {https://doi.org/10.1063/1.3630964}
  {\bibfield  {journal} {\bibinfo  {journal} {J. Appl. Phys.}\ }\textbf
  {\bibinfo {volume} {110}},\ \bibinfo {pages} {053714} (\bibinfo {year}
  {2011})}\BibitemShut {NoStop}%
\bibitem [{\citenamefont {Sezen}\ \emph {et~al.}(2014)\citenamefont {Sezen},
  \citenamefont {Buchholz}, \citenamefont {Nefedov}, \citenamefont {Heissler},
  \citenamefont {Di~Valentin},\ and\ \citenamefont {W{\"o}ll}}]{Sezen2014}%
  \BibitemOpen
  \bibfield  {author} {\bibinfo {author} {\bibfnamefont {H.}~\bibnamefont
  {Sezen}}, \bibinfo {author} {\bibfnamefont {M.}~\bibnamefont {Buchholz}},
  \bibinfo {author} {\bibfnamefont {A.}~\bibnamefont {Nefedov}}, \bibinfo
  {author} {\bibfnamefont {S.}~\bibnamefont {Heissler}}, \bibinfo {author}
  {\bibfnamefont {C.}~\bibnamefont {Di~Valentin}},\ and\ \bibinfo {author}
  {\bibfnamefont {C.}~\bibnamefont {W{\"o}ll}},\ }\bibfield  {title} {\bibinfo
  {title} {Probing electrons in {T}i{O}$_2$ polaronic trap states by
  {IR}-absorption: {E}vidence for the existence of hydrogenic states},\
  }\bibfield  {journal} {\bibinfo  {journal} {Sci. Rep.}\ }\textbf {\bibinfo
  {volume} {4}},\ \href {https://doi.org/10.1038/srep03808} {10.1038/srep03808}
  (\bibinfo {year} {2014})\BibitemShut {NoStop}%
\bibitem [{\citenamefont {Lavrov~E.}(2015)}]{Lavrov2015}%
  \BibitemOpen
  \bibfield  {author} {\bibinfo {author} {\bibfnamefont {V.}~\bibnamefont
  {Lavrov~E.}},\ }\bibfield  {title} {\bibinfo {title} {Infrared absorption on
  hydrogen in anatase {T}i{O}$_2$},\ }\href
  {https://doi.org/10.1002/pssa.201431814} {\bibfield  {journal} {\bibinfo
  {journal} {Phys. Status Solidi A}\ }\textbf {\bibinfo {volume} {212}},\
  \bibinfo {pages} {1494} (\bibinfo {year} {2015})}\BibitemShut {NoStop}%
\bibitem [{\citenamefont {Lavrov}(2016)}]{Lavrov2016}%
  \BibitemOpen
  \bibfield  {author} {\bibinfo {author} {\bibfnamefont {E.~V.}\ \bibnamefont
  {Lavrov}},\ }\bibfield  {title} {\bibinfo {title} {Hydrogen donor in anatase
  ${\mathrm{tio}}_{2}$},\ }\href {https://doi.org/10.1103/PhysRevB.93.045204}
  {\bibfield  {journal} {\bibinfo  {journal} {Phys. Rev. B}\ }\textbf {\bibinfo
  {volume} {93}},\ \bibinfo {pages} {045204} (\bibinfo {year}
  {2016})}\BibitemShut {NoStop}%
\bibitem [{\citenamefont {Yu}\ \emph {et~al.}(2013)\citenamefont {Yu},
  \citenamefont {Kim},\ and\ \citenamefont {Kim}}]{Yu2013}%
  \BibitemOpen
  \bibfield  {author} {\bibinfo {author} {\bibfnamefont {X.}~\bibnamefont
  {Yu}}, \bibinfo {author} {\bibfnamefont {B.}~\bibnamefont {Kim}},\ and\
  \bibinfo {author} {\bibfnamefont {Y.~K.}\ \bibnamefont {Kim}},\ }\bibfield
  {title} {\bibinfo {title} {Highly {E}nhanced {P}hotoactivity of {A}natase
  {T}i{O}$_2$ {N}anocrystals by {C}ontrolled {H}ydrogenation-{I}nduced
  {S}urface {D}efects},\ }\href {https://doi.org/10.1021/cs4005776} {\bibfield
  {journal} {\bibinfo  {journal} {ACS Catal.}\ }\textbf {\bibinfo {volume}
  {3}},\ \bibinfo {pages} {2479} (\bibinfo {year} {2013})}\BibitemShut
  {NoStop}%
\bibitem [{\citenamefont {Johnston}\ \emph {et~al.}(2017)\citenamefont
  {Johnston}, \citenamefont {Schell}, \citenamefont {Correia}, \citenamefont
  {Deicher}, \citenamefont {Gunnlaugsson}, \citenamefont {Fenta}, \citenamefont
  {David-Bosne}, \citenamefont {Costa},\ and\ \citenamefont
  {Lupascu}}]{Johnston2017}%
  \BibitemOpen
  \bibfield  {author} {\bibinfo {author} {\bibfnamefont {K.}~\bibnamefont
  {Johnston}}, \bibinfo {author} {\bibfnamefont {J.}~\bibnamefont {Schell}},
  \bibinfo {author} {\bibfnamefont {J.~G.}\ \bibnamefont {Correia}}, \bibinfo
  {author} {\bibfnamefont {M.}~\bibnamefont {Deicher}}, \bibinfo {author}
  {\bibfnamefont {H.~P.}\ \bibnamefont {Gunnlaugsson}}, \bibinfo {author}
  {\bibfnamefont {A.~S.}\ \bibnamefont {Fenta}}, \bibinfo {author}
  {\bibfnamefont {E.}~\bibnamefont {David-Bosne}}, \bibinfo {author}
  {\bibfnamefont {A.~R.~G.}\ \bibnamefont {Costa}},\ and\ \bibinfo {author}
  {\bibfnamefont {D.~C.}\ \bibnamefont {Lupascu}},\ }\bibfield  {title}
  {\bibinfo {title} {The solid state physics programme at {ISOLDE}: Recent
  developments and perspectives},\ }\href
  {https://doi.org/10.1088/1361-6471/aa81ac} {\bibfield  {journal} {\bibinfo
  {journal} {J. Phys. G: Nucl. Part. Phys.}\ }\textbf {\bibinfo {volume}
  {44}},\ \bibinfo {pages} {104001} (\bibinfo {year} {2017})}\BibitemShut
  {NoStop}%
\bibitem [{\citenamefont {Kresse}\ and\ \citenamefont
  {Furthm{\"u}ller}(1996)}]{KRESSE1996}%
  \BibitemOpen
  \bibfield  {author} {\bibinfo {author} {\bibfnamefont {G.}~\bibnamefont
  {Kresse}}\ and\ \bibinfo {author} {\bibfnamefont {J.}~\bibnamefont
  {Furthm{\"u}ller}},\ }\bibfield  {title} {\bibinfo {title} {Efficiency of
  ab-initio total energy calculations for metals and semiconductors using a
  plane-wave basis set},\ }\href {https://doi.org/10.1016/0927-0256(96)00008-0}
  {\bibfield  {journal} {\bibinfo  {journal} {Comput. Mater. Sci.}\ }\textbf
  {\bibinfo {volume} {6}},\ \bibinfo {pages} {15 } (\bibinfo {year}
  {1996})}\BibitemShut {NoStop}%
\bibitem [{\citenamefont {Blaha}\ \emph {et~al.}(2008)\citenamefont {Blaha},
  \citenamefont {Schwarz}, \citenamefont {Madsen}, \citenamefont {Kvasnicka},
  \citenamefont {Luitz},\ and\ \citenamefont {Schwarz}}]{blaha2008}%
  \BibitemOpen
  \bibfield  {author} {\bibinfo {author} {\bibfnamefont {P.}~\bibnamefont
  {Blaha}}, \bibinfo {author} {\bibfnamefont {K.}~\bibnamefont {Schwarz}},
  \bibinfo {author} {\bibfnamefont {G.}~\bibnamefont {Madsen}}, \bibinfo
  {author} {\bibfnamefont {D.}~\bibnamefont {Kvasnicka}}, \bibinfo {author}
  {\bibfnamefont {J.}~\bibnamefont {Luitz}},\ and\ \bibinfo {author}
  {\bibfnamefont {K.}~\bibnamefont {Schwarz}},\ }\bibfield  {title} {\bibinfo
  {title} {An {A}ugmented {P}lane {W}ave plus local {O}rbitals {P}rogram for
  {C}alculating {C}rystal {P}roperties: {W}ien2{K} {U}ser's {G}uide
  ({W}ien:{T}echnische {U}niversitat {W}ien)},\ }\href@noop {} {\bibfield
  {journal} {\bibinfo  {journal} {TU Wien}\ } (\bibinfo {year}
  {2008})}\BibitemShut {NoStop}%
\bibitem [{\citenamefont {Vlaic}\ \emph {et~al.}(2015)\citenamefont {Vlaic},
  \citenamefont {Ivanov}, \citenamefont {Peipmann}, \citenamefont {Eisenhardt},
  \citenamefont {Himmerlich}, \citenamefont {Krischok},\ and\ \citenamefont
  {Bund}}]{Vlaic2015}%
  \BibitemOpen
  \bibfield  {author} {\bibinfo {author} {\bibfnamefont {C.~A.}\ \bibnamefont
  {Vlaic}}, \bibinfo {author} {\bibfnamefont {S.}~\bibnamefont {Ivanov}},
  \bibinfo {author} {\bibfnamefont {R.}~\bibnamefont {Peipmann}}, \bibinfo
  {author} {\bibfnamefont {A.}~\bibnamefont {Eisenhardt}}, \bibinfo {author}
  {\bibfnamefont {M.}~\bibnamefont {Himmerlich}}, \bibinfo {author}
  {\bibfnamefont {S.}~\bibnamefont {Krischok}},\ and\ \bibinfo {author}
  {\bibfnamefont {A.}~\bibnamefont {Bund}},\ }\bibfield  {title} {\bibinfo
  {title} {Electrochemical lithiation of thin silicon based layers
  potentiostatically deposited from ionic liquid},\ }\href
  {https://doi.org/10.1016/j.electacta.2015.03.216} {\bibfield  {journal}
  {\bibinfo  {journal} {Electrochim. Acta}\ }\textbf {\bibinfo {volume}
  {168}},\ \bibinfo {pages} {403 } (\bibinfo {year} {2015})}\BibitemShut
  {NoStop}%
\bibitem [{\citenamefont {Borge}\ and\ \citenamefont
  {Jonson}(2017)}]{Borge2017}%
  \BibitemOpen
  \bibfield  {author} {\bibinfo {author} {\bibfnamefont {M.~J.~G.}\
  \bibnamefont {Borge}}\ and\ \bibinfo {author} {\bibfnamefont
  {B.}~\bibnamefont {Jonson}},\ }\bibfield  {title} {\bibinfo {title} {{ISOLDE}
  past, present and future},\ }\href {https://doi.org/10.1088/1361-6471/aa5f03}
  {\bibfield  {journal} {\bibinfo  {journal} {J. Phys. G: Nucl. Part. Phys.}\
  }\textbf {\bibinfo {volume} {44}},\ \bibinfo {pages} {044011} (\bibinfo
  {year} {2017})}\BibitemShut {NoStop}%
\bibitem [{\citenamefont {Catherall}\ \emph {et~al.}(2017)\citenamefont
  {Catherall}, \citenamefont {Andreazza}, \citenamefont {Breitenfeldt},
  \citenamefont {Dorsival}, \citenamefont {Focker}, \citenamefont {Gharsa},
  \citenamefont {Giles~T}, \citenamefont {Grenard}, \citenamefont {Locci},
  \citenamefont {Martins}, \citenamefont {Marzari}, \citenamefont {Schipper},
  \citenamefont {Shornikov},\ and\ \citenamefont {Stora}}]{Catherall2017}%
  \BibitemOpen
  \bibfield  {author} {\bibinfo {author} {\bibfnamefont {R.}~\bibnamefont
  {Catherall}}, \bibinfo {author} {\bibfnamefont {W.}~\bibnamefont
  {Andreazza}}, \bibinfo {author} {\bibfnamefont {M.}~\bibnamefont
  {Breitenfeldt}}, \bibinfo {author} {\bibfnamefont {A.}~\bibnamefont
  {Dorsival}}, \bibinfo {author} {\bibfnamefont {G.~J.}\ \bibnamefont
  {Focker}}, \bibinfo {author} {\bibfnamefont {T.~P.}\ \bibnamefont {Gharsa}},
  \bibinfo {author} {\bibfnamefont {J.}~\bibnamefont {Giles~T}}, \bibinfo
  {author} {\bibfnamefont {J.-L.}\ \bibnamefont {Grenard}}, \bibinfo {author}
  {\bibfnamefont {F.}~\bibnamefont {Locci}}, \bibinfo {author} {\bibfnamefont
  {P.}~\bibnamefont {Martins}}, \bibinfo {author} {\bibfnamefont
  {S.}~\bibnamefont {Marzari}}, \bibinfo {author} {\bibfnamefont
  {J.}~\bibnamefont {Schipper}}, \bibinfo {author} {\bibfnamefont
  {A.}~\bibnamefont {Shornikov}},\ and\ \bibinfo {author} {\bibfnamefont
  {T.}~\bibnamefont {Stora}},\ }\bibfield  {title} {\bibinfo {title} {The
  {ISOLDE} facility},\ }\href {https://doi.org/10.1088/1361-6471/aa7eba}
  {\bibfield  {journal} {\bibinfo  {journal} {J. Phys. G: Nucl. Part. Phys.}\
  }\textbf {\bibinfo {volume} {44}},\ \bibinfo {pages} {094002} (\bibinfo
  {year} {2017})}\BibitemShut {NoStop}%
\bibitem [{\citenamefont {Fedoseyev}\ \emph {et~al.}(2000)\citenamefont
  {Fedoseyev}, \citenamefont {Huber}, \citenamefont {K{\"o}ster}, \citenamefont
  {Lettry}, \citenamefont {Mishin}, \citenamefont {Ravn},\ and\ \citenamefont
  {Sebastian}}]{Fedoseyev2000}%
  \BibitemOpen
  \bibfield  {author} {\bibinfo {author} {\bibfnamefont {V.}~\bibnamefont
  {Fedoseyev}}, \bibinfo {author} {\bibfnamefont {G.}~\bibnamefont {Huber}},
  \bibinfo {author} {\bibfnamefont {U.}~\bibnamefont {K{\"o}ster}}, \bibinfo
  {author} {\bibfnamefont {J.}~\bibnamefont {Lettry}}, \bibinfo {author}
  {\bibfnamefont {V.}~\bibnamefont {Mishin}}, \bibinfo {author} {\bibfnamefont
  {H.}~\bibnamefont {Ravn}},\ and\ \bibinfo {author} {\bibfnamefont
  {V.}~\bibnamefont {Sebastian}},\ }\bibfield  {title} {\bibinfo {title} {The
  {ISOLDE} laser ion source for exotic nuclei},\ }\href
  {https://doi.org/10.1023/A:1012609515865} {\bibfield  {journal} {\bibinfo
  {journal} {Hyperfine Interact.}\ }\textbf {\bibinfo {volume} {127}},\
  \bibinfo {pages} {409} (\bibinfo {year} {2000})}\BibitemShut {NoStop}%
\bibitem [{\citenamefont {Ziegler}\ \emph {et~al.}(2010)\citenamefont
  {Ziegler}, \citenamefont {Ziegler},\ and\ \citenamefont
  {Biersack}}]{Ziegler2010}%
  \BibitemOpen
  \bibfield  {author} {\bibinfo {author} {\bibfnamefont {J.~F.}\ \bibnamefont
  {Ziegler}}, \bibinfo {author} {\bibfnamefont {M.~D.}\ \bibnamefont
  {Ziegler}},\ and\ \bibinfo {author} {\bibfnamefont {J.~P.}\ \bibnamefont
  {Biersack}},\ }\bibfield  {title} {\bibinfo {title} {{SRIM} - {T}he stopping
  and range of ions in matter (2010)},\ }\href
  {https://doi.org/10.1016/j.nimb.2010.02.091} {\bibfield  {journal} {\bibinfo
  {journal} {Nucl. Instrum. Methods Phys. Res., Sect. B}\ }\textbf {\bibinfo
  {volume} {268}},\ \bibinfo {pages} {1818} (\bibinfo {year}
  {2010})}\BibitemShut {NoStop}%
\bibitem [{\citenamefont {Sun}\ \emph {et~al.}(2011)\citenamefont {Sun},
  \citenamefont {Jia}, \citenamefont {Yang}, \citenamefont {Yang},
  \citenamefont {Yao}, \citenamefont {Lu}, \citenamefont {Selloni},\ and\
  \citenamefont {Smith}}]{Sun2011}%
  \BibitemOpen
  \bibfield  {author} {\bibinfo {author} {\bibfnamefont {C.}~\bibnamefont
  {Sun}}, \bibinfo {author} {\bibfnamefont {Y.}~\bibnamefont {Jia}}, \bibinfo
  {author} {\bibfnamefont {X.-H.}\ \bibnamefont {Yang}}, \bibinfo {author}
  {\bibfnamefont {H.-G.}\ \bibnamefont {Yang}}, \bibinfo {author}
  {\bibfnamefont {X.}~\bibnamefont {Yao}}, \bibinfo {author} {\bibfnamefont
  {G.~Q.~M.}\ \bibnamefont {Lu}}, \bibinfo {author} {\bibfnamefont
  {A.}~\bibnamefont {Selloni}},\ and\ \bibinfo {author} {\bibfnamefont {S.~C.}\
  \bibnamefont {Smith}},\ }\bibfield  {title} {\bibinfo {title} {Hydrogen
  {I}ncorporation and {S}torage in {W}ell-{D}efined {N}anocrystals of {A}natase
  {T}itanium {D}ioxide},\ }\href {https://doi.org/10.1021/jp210472p} {\bibfield
   {journal} {\bibinfo  {journal} {J. Phys. Chem. C}\ }\textbf {\bibinfo
  {volume} {115}},\ \bibinfo {pages} {25590} (\bibinfo {year}
  {2011})}\BibitemShut {NoStop}%
\bibitem [{\citenamefont {Diebold}(2003)}]{Diebold2003}%
  \BibitemOpen
  \bibfield  {author} {\bibinfo {author} {\bibfnamefont {U.}~\bibnamefont
  {Diebold}},\ }\bibfield  {title} {\bibinfo {title} {The surface science of
  titanium dioxide},\ }\href {https://doi.org/10.1016/S0167-5729(02)00100-0}
  {\bibfield  {journal} {\bibinfo  {journal} {Surf. Sci. Rep.}\ }\textbf
  {\bibinfo {volume} {48}},\ \bibinfo {pages} {53} (\bibinfo {year}
  {2003})}\BibitemShut {NoStop}%
\bibitem [{\citenamefont {Naldoni}\ \emph {et~al.}(2012)\citenamefont
  {Naldoni}, \citenamefont {Allieta}, \citenamefont {Santangelo}, \citenamefont
  {Marelli}, \citenamefont {Fabbri}, \citenamefont {Cappelli}, \citenamefont
  {Bianchi}, \citenamefont {Psaro},\ and\ \citenamefont
  {Dal~Santo}}]{Naldoni2012}%
  \BibitemOpen
  \bibfield  {author} {\bibinfo {author} {\bibfnamefont {A.}~\bibnamefont
  {Naldoni}}, \bibinfo {author} {\bibfnamefont {M.}~\bibnamefont {Allieta}},
  \bibinfo {author} {\bibfnamefont {S.}~\bibnamefont {Santangelo}}, \bibinfo
  {author} {\bibfnamefont {M.}~\bibnamefont {Marelli}}, \bibinfo {author}
  {\bibfnamefont {F.}~\bibnamefont {Fabbri}}, \bibinfo {author} {\bibfnamefont
  {S.}~\bibnamefont {Cappelli}}, \bibinfo {author} {\bibfnamefont {C.~L.}\
  \bibnamefont {Bianchi}}, \bibinfo {author} {\bibfnamefont {R.}~\bibnamefont
  {Psaro}},\ and\ \bibinfo {author} {\bibfnamefont {V.}~\bibnamefont
  {Dal~Santo}},\ }\bibfield  {title} {\bibinfo {title} {Effect of {N}ature and
  {L}ocation of {D}efects on {B}andgap {N}arrowing in {B}lack {T}i{O}$_2$
  {N}anoparticles},\ }\href {https://doi.org/10.1021/ja3012676} {\bibfield
  {journal} {\bibinfo  {journal} {J. Am. Chem. Soc.}\ }\textbf {\bibinfo
  {volume} {134}},\ \bibinfo {pages} {7600} (\bibinfo {year}
  {2012})}\BibitemShut {NoStop}%
\bibitem [{\citenamefont {Tauc}\ \emph {et~al.}(2006)\citenamefont {Tauc},
  \citenamefont {Grigorovici},\ and\ \citenamefont {Vancu}}]{Tauc2006}%
  \BibitemOpen
  \bibfield  {author} {\bibinfo {author} {\bibfnamefont {J.}~\bibnamefont
  {Tauc}}, \bibinfo {author} {\bibfnamefont {R.}~\bibnamefont {Grigorovici}},\
  and\ \bibinfo {author} {\bibfnamefont {A.}~\bibnamefont {Vancu}},\ }\bibfield
   {title} {\bibinfo {title} {Optical {P}roperties and {E}lectronic {S}tructure
  of {A}morphous {G}ermanium},\ }\href
  {https://doi.org/10.1002/pssb.19660150224} {\bibfield  {journal} {\bibinfo
  {journal} {phys. stat. sol. (b)}\ }\textbf {\bibinfo {volume} {15}},\
  \bibinfo {pages} {627} (\bibinfo {year} {2006})}\BibitemShut {NoStop}%
\bibitem [{\citenamefont {He}\ \emph {et~al.}(2013)\citenamefont {He},
  \citenamefont {Yang}, \citenamefont {Wang}, \citenamefont {Luo},\ and\
  \citenamefont {Chen}}]{He2013}%
  \BibitemOpen
  \bibfield  {author} {\bibinfo {author} {\bibfnamefont {H.}~\bibnamefont
  {He}}, \bibinfo {author} {\bibfnamefont {K.}~\bibnamefont {Yang}}, \bibinfo
  {author} {\bibfnamefont {N.}~\bibnamefont {Wang}}, \bibinfo {author}
  {\bibfnamefont {F.}~\bibnamefont {Luo}},\ and\ \bibinfo {author}
  {\bibfnamefont {H.}~\bibnamefont {Chen}},\ }\bibfield  {title} {\bibinfo
  {title} {Hydrogenated {T}i{O}$_2$ film for enhancing photovoltaic properties
  of solar cells and self-sensitized effect},\ }\href
  {https://doi.org/10.1063/1.4832783} {\bibfield  {journal} {\bibinfo
  {journal} {J. Appl. Phys.}\ }\textbf {\bibinfo {volume} {114}},\ \bibinfo
  {pages} {213505} (\bibinfo {year} {2013})}\BibitemShut {NoStop}%
\bibitem [{\citenamefont {Mohammadizadeh}\ \emph {et~al.}(2015)\citenamefont
  {Mohammadizadeh}, \citenamefont {Bagheri}, \citenamefont {Aghabagheri},\ and\
  \citenamefont {Abdi}}]{Mohammadizadeh2015}%
  \BibitemOpen
  \bibfield  {author} {\bibinfo {author} {\bibfnamefont {M.~R.}\ \bibnamefont
  {Mohammadizadeh}}, \bibinfo {author} {\bibfnamefont {M.}~\bibnamefont
  {Bagheri}}, \bibinfo {author} {\bibfnamefont {S.}~\bibnamefont
  {Aghabagheri}},\ and\ \bibinfo {author} {\bibfnamefont {Y.}~\bibnamefont
  {Abdi}},\ }\bibfield  {title} {\bibinfo {title} {Photocatalytic activity of
  {T}i{O}$_2$ thin films by hydrogen {DC} plasma},\ }\href
  {https://doi.org/10.1016/j.apsusc.2015.03.196} {\bibfield  {journal}
  {\bibinfo  {journal} {Appl. Surf. Sci.}\ }\textbf {\bibinfo {volume} {350}},\
  \bibinfo {pages} {43} (\bibinfo {year} {2015})}\BibitemShut {NoStop}%
\bibitem [{\citenamefont {Noerenberg}\ and\ \citenamefont
  {Briggs}(1998)}]{Noerenberg1998}%
  \BibitemOpen
  \bibfield  {author} {\bibinfo {author} {\bibfnamefont {H.}~\bibnamefont
  {Noerenberg}}\ and\ \bibinfo {author} {\bibfnamefont {G.}~\bibnamefont
  {Briggs}},\ }\bibfield  {title} {\bibinfo {title} {Surface structure of the
  most oxygen deficient {M}agn\'eli phase – an {STM} study of
  {T}i$_4${O}$_7$},\ }\href {https://doi.org/10.1016/S0039-6028(97)01000-5}
  {\bibfield  {journal} {\bibinfo  {journal} {Surf. Sci.}\ }\textbf {\bibinfo
  {volume} {402-404}},\ \bibinfo {pages} {738 } (\bibinfo {year}
  {1998})}\BibitemShut {NoStop}%
\bibitem [{\citenamefont {Cronemeyer}(1959)}]{Cronemeyer1959}%
  \BibitemOpen
  \bibfield  {author} {\bibinfo {author} {\bibfnamefont {D.~C.}\ \bibnamefont
  {Cronemeyer}},\ }\bibfield  {title} {\bibinfo {title} {Infrared {A}bsorption
  of {R}educed {R}utile {T}i${\mathrm{o}}_{2}$ {S}ingle {C}rystals},\ }\href
  {https://doi.org/doi/10.1103/PhysRev.113.1222} {\bibfield  {journal}
  {\bibinfo  {journal} {Phys. Rev.}\ }\textbf {\bibinfo {volume} {113}},\
  \bibinfo {pages} {1222} (\bibinfo {year} {1959})}\BibitemShut {NoStop}%
\bibitem [{\citenamefont {Ivanov}\ \emph {et~al.}(2016)\citenamefont {Ivanov},
  \citenamefont {Barylyak}, \citenamefont {Besaha}, \citenamefont {Dimitrova},
  \citenamefont {Krischok}, \citenamefont {Bund},\ and\ \citenamefont
  {Bobitski}}]{Ivanov2016}%
  \BibitemOpen
  \bibfield  {author} {\bibinfo {author} {\bibfnamefont {S.}~\bibnamefont
  {Ivanov}}, \bibinfo {author} {\bibfnamefont {A.}~\bibnamefont {Barylyak}},
  \bibinfo {author} {\bibfnamefont {K.}~\bibnamefont {Besaha}}, \bibinfo
  {author} {\bibfnamefont {A.}~\bibnamefont {Dimitrova}}, \bibinfo {author}
  {\bibfnamefont {S.}~\bibnamefont {Krischok}}, \bibinfo {author}
  {\bibfnamefont {A.}~\bibnamefont {Bund}},\ and\ \bibinfo {author}
  {\bibfnamefont {J.}~\bibnamefont {Bobitski}},\ }\bibfield  {title} {\bibinfo
  {title} {Enhanced lithium ion storage in {T}i{O}$_2$ nanoparticles, induced
  by sulphur and carbon co-doping},\ }\href
  {https://doi.org/10.1016/j.jpowsour.2016.06.116} {\bibfield  {journal}
  {\bibinfo  {journal} {J. Power Sources}\ }\textbf {\bibinfo {volume} {326}},\
  \bibinfo {pages} {270} (\bibinfo {year} {2016})}\BibitemShut {NoStop}%
\bibitem [{\citenamefont {Wang}\ \emph {et~al.}(2011)\citenamefont {Wang},
  \citenamefont {Wang}, \citenamefont {Ling}, \citenamefont {Tang},
  \citenamefont {Yang}, \citenamefont {Fitzmorris}, \citenamefont {Wang},
  \citenamefont {Zhang},\ and\ \citenamefont {Li}}]{Wang2011}%
  \BibitemOpen
  \bibfield  {author} {\bibinfo {author} {\bibfnamefont {G.}~\bibnamefont
  {Wang}}, \bibinfo {author} {\bibfnamefont {H.}~\bibnamefont {Wang}}, \bibinfo
  {author} {\bibfnamefont {Y.}~\bibnamefont {Ling}}, \bibinfo {author}
  {\bibfnamefont {Y.}~\bibnamefont {Tang}}, \bibinfo {author} {\bibfnamefont
  {X.}~\bibnamefont {Yang}}, \bibinfo {author} {\bibfnamefont {R.~C.}\
  \bibnamefont {Fitzmorris}}, \bibinfo {author} {\bibfnamefont
  {C.}~\bibnamefont {Wang}}, \bibinfo {author} {\bibfnamefont {J.~Z.}\
  \bibnamefont {Zhang}},\ and\ \bibinfo {author} {\bibfnamefont
  {Y.}~\bibnamefont {Li}},\ }\bibfield  {title} {\bibinfo {title}
  {Hydrogen-{T}reated {T}io$_2$ {N}anowire {A}rrays for {P}hotoelectrochemical
  {W}ater {S}plitting},\ }\href {https://doi.org/10.1021/nl201766h} {\bibfield
  {journal} {\bibinfo  {journal} {Nano Lett.}\ }\textbf {\bibinfo {volume}
  {11}},\ \bibinfo {pages} {3026} (\bibinfo {year} {2011})}\BibitemShut
  {NoStop}%
\bibitem [{\citenamefont {Yan}\ \emph {et~al.}(2014)\citenamefont {Yan},
  \citenamefont {Han}, \citenamefont {Konkin}, \citenamefont {Koppe},
  \citenamefont {Wang}, \citenamefont {Andreu}, \citenamefont {Chen},
  \citenamefont {Vetter}, \citenamefont {Morante},\ and\ \citenamefont
  {Schaaf}}]{Yan2014}%
  \BibitemOpen
  \bibfield  {author} {\bibinfo {author} {\bibfnamefont {Y.}~\bibnamefont
  {Yan}}, \bibinfo {author} {\bibfnamefont {M.}~\bibnamefont {Han}}, \bibinfo
  {author} {\bibfnamefont {A.}~\bibnamefont {Konkin}}, \bibinfo {author}
  {\bibfnamefont {T.}~\bibnamefont {Koppe}}, \bibinfo {author} {\bibfnamefont
  {D.}~\bibnamefont {Wang}}, \bibinfo {author} {\bibfnamefont {T.}~\bibnamefont
  {Andreu}}, \bibinfo {author} {\bibfnamefont {G.}~\bibnamefont {Chen}},
  \bibinfo {author} {\bibfnamefont {U.}~\bibnamefont {Vetter}}, \bibinfo
  {author} {\bibfnamefont {J.~R.}\ \bibnamefont {Morante}},\ and\ \bibinfo
  {author} {\bibfnamefont {P.}~\bibnamefont {Schaaf}},\ }\bibfield  {title}
  {\bibinfo {title} {Slightly hydrogenated {T}i{O}$_2$ with enhanced
  photocatalytic performance},\ }\href {https://doi.org/10.1039/C4TA02192D}
  {\bibfield  {journal} {\bibinfo  {journal} {J. Mater. Chem. A}\ }\textbf
  {\bibinfo {volume} {2}},\ \bibinfo {pages} {12708} (\bibinfo {year}
  {2014})}\BibitemShut {NoStop}%
\bibitem [{\citenamefont {Panomsuwan}\ \emph {et~al.}(2015)\citenamefont
  {Panomsuwan}, \citenamefont {Watthanaphanit}, \citenamefont {Ishizaki},\ and\
  \citenamefont {Saito}}]{Panomsuwan2015}%
  \BibitemOpen
  \bibfield  {author} {\bibinfo {author} {\bibfnamefont {G.}~\bibnamefont
  {Panomsuwan}}, \bibinfo {author} {\bibfnamefont {A.}~\bibnamefont
  {Watthanaphanit}}, \bibinfo {author} {\bibfnamefont {T.}~\bibnamefont
  {Ishizaki}},\ and\ \bibinfo {author} {\bibfnamefont {N.}~\bibnamefont
  {Saito}},\ }\bibfield  {title} {\bibinfo {title} {Water-plasma-assisted
  synthesis of black titania spheres with efficient visible-light
  photocatalytic activity},\ }\href {https://doi.org/10.1039/C5CP00171D}
  {\bibfield  {journal} {\bibinfo  {journal} {Phys. Chem. Chem. Phys.}\
  }\textbf {\bibinfo {volume} {17}},\ \bibinfo {pages} {13794} (\bibinfo {year}
  {2015})}\BibitemShut {NoStop}%
\bibitem [{\citenamefont {Singh}\ \emph {et~al.}(2016)\citenamefont {Singh},
  \citenamefont {Kodan}, \citenamefont {Mehta}, \citenamefont {Dey},\ and\
  \citenamefont {Krishnamurthy}}]{Singh2016}%
  \BibitemOpen
  \bibfield  {author} {\bibinfo {author} {\bibfnamefont {A.~P.}\ \bibnamefont
  {Singh}}, \bibinfo {author} {\bibfnamefont {N.}~\bibnamefont {Kodan}},
  \bibinfo {author} {\bibfnamefont {B.~R.}\ \bibnamefont {Mehta}}, \bibinfo
  {author} {\bibfnamefont {A.}~\bibnamefont {Dey}},\ and\ \bibinfo {author}
  {\bibfnamefont {S.}~\bibnamefont {Krishnamurthy}},\ }\bibfield  {title}
  {\bibinfo {title} {In-situ plasma hydrogenated {T}i{O}$_2$ thin films for
  enhanced photoelectrochemical properties},\ }\href
  {https://doi.org/10.1016/j.jpowsour.2016.06.116} {\bibfield  {journal}
  {\bibinfo  {journal} {Mater. Res. Bull.}\ }\textbf {\bibinfo {volume} {76}},\
  \bibinfo {pages} {284} (\bibinfo {year} {2016})}\BibitemShut {NoStop}%
\bibitem [{\citenamefont {Fan}\ \emph {et~al.}(2015)\citenamefont {Fan},
  \citenamefont {Chen}, \citenamefont {Wang}, \citenamefont {Fu}, \citenamefont
  {Ren}, \citenamefont {Qian},\ and\ \citenamefont {Wang}}]{Fan2015}%
  \BibitemOpen
  \bibfield  {author} {\bibinfo {author} {\bibfnamefont {C.}~\bibnamefont
  {Fan}}, \bibinfo {author} {\bibfnamefont {C.}~\bibnamefont {Chen}}, \bibinfo
  {author} {\bibfnamefont {J.}~\bibnamefont {Wang}}, \bibinfo {author}
  {\bibfnamefont {X.}~\bibnamefont {Fu}}, \bibinfo {author} {\bibfnamefont
  {Z.}~\bibnamefont {Ren}}, \bibinfo {author} {\bibfnamefont {G.}~\bibnamefont
  {Qian}},\ and\ \bibinfo {author} {\bibfnamefont {Z.}~\bibnamefont {Wang}},\
  }\bibfield  {title} {\bibinfo {title} {Black {H}ydroxylated {T}itanium
  {D}ioxide {P}repared via {U}ltrasonication with {E}nhanced {P}hotocatalytic
  {A}ctivity},\ }\href {https://doi.org/10.1038/srep11712} {\bibfield
  {journal} {\bibinfo  {journal} {Sci. Rep.}\ }\textbf {\bibinfo {volume}
  {5}},\ \bibinfo {pages} {11712} (\bibinfo {year} {2015})}\BibitemShut
  {NoStop}%
\bibitem [{\citenamefont {McCafferty}\ and\ \citenamefont
  {Wightman~J.}(1998)}]{McCafferty1998}%
  \BibitemOpen
  \bibfield  {author} {\bibinfo {author} {\bibfnamefont {E.}~\bibnamefont
  {McCafferty}}\ and\ \bibinfo {author} {\bibfnamefont {P.}~\bibnamefont
  {Wightman~J.}},\ }\bibfield  {title} {\bibinfo {title} {Determination of the
  concentration of surface hydroxyl groups on metal oxide films by a
  quantitative {XPS} method},\ }\href
  {https://doi.org/10.1002/(sici)1096-9918(199807)26:8<549::aid-sia396>3.0.co;2-q}
  {\bibfield  {journal} {\bibinfo  {journal} {Surf. Interface Anal.}\ }\textbf
  {\bibinfo {volume} {26}},\ \bibinfo {pages} {549} (\bibinfo {year}
  {1998})}\BibitemShut {NoStop}%
\bibitem [{\citenamefont {Zhou}\ \emph {et~al.}(2013)\citenamefont {Zhou},
  \citenamefont {Chongyin}, \citenamefont {Tianquan}, \citenamefont {Hao},
  \citenamefont {Ping}, \citenamefont {Dongyun}, \citenamefont {Fangfang},
  \citenamefont {Fuqiang}, \citenamefont {Jianhua}, \citenamefont {Xiaoming},\
  and\ \citenamefont {Mianheng}}]{Zhou2013}%
  \BibitemOpen
  \bibfield  {author} {\bibinfo {author} {\bibfnamefont {W.}~\bibnamefont
  {Zhou}}, \bibinfo {author} {\bibfnamefont {Y.}~\bibnamefont {Chongyin}},
  \bibinfo {author} {\bibfnamefont {L.}~\bibnamefont {Tianquan}}, \bibinfo
  {author} {\bibfnamefont {Y.}~\bibnamefont {Hao}}, \bibinfo {author}
  {\bibfnamefont {C.}~\bibnamefont {Ping}}, \bibinfo {author} {\bibfnamefont
  {W.}~\bibnamefont {Dongyun}}, \bibinfo {author} {\bibfnamefont
  {X.}~\bibnamefont {Fangfang}}, \bibinfo {author} {\bibfnamefont
  {H.}~\bibnamefont {Fuqiang}}, \bibinfo {author} {\bibfnamefont
  {L.}~\bibnamefont {Jianhua}}, \bibinfo {author} {\bibfnamefont
  {X.}~\bibnamefont {Xiaoming}},\ and\ \bibinfo {author} {\bibfnamefont
  {J.}~\bibnamefont {Mianheng}},\ }\bibfield  {title} {\bibinfo {title}
  {H-{D}oped {B}lack {T}itania with {V}ery {H}igh {S}olar {A}bsorption and
  {E}xcellent {P}hotocatalysis {E}nhanced by {L}ocalized {S}urface {P}lasmon
  {R}esonance},\ }\href {https://doi.org/10.1002/adfm.201300486} {\bibfield
  {journal} {\bibinfo  {journal} {Adv. Funct. Mater.}\ }\textbf {\bibinfo
  {volume} {23}},\ \bibinfo {pages} {5444} (\bibinfo {year}
  {2013})}\BibitemShut {NoStop}%
\bibitem [{\citenamefont {Amano}\ \emph {et~al.}(2016)\citenamefont {Amano},
  \citenamefont {Nakata}, \citenamefont {Yamamoto},\ and\ \citenamefont
  {Tanaka}}]{Amano2016}%
  \BibitemOpen
  \bibfield  {author} {\bibinfo {author} {\bibfnamefont {F.}~\bibnamefont
  {Amano}}, \bibinfo {author} {\bibfnamefont {M.}~\bibnamefont {Nakata}},
  \bibinfo {author} {\bibfnamefont {A.}~\bibnamefont {Yamamoto}},\ and\
  \bibinfo {author} {\bibfnamefont {T.}~\bibnamefont {Tanaka}},\ }\bibfield
  {title} {\bibinfo {title} {Effect of {T}i$^{3+}$ {I}ons and {C}onduction
  {B}and {E}lectrons on {P}hotocatalytic and {P}hotoelectrochemical {A}ctivity
  of {R}utile {T}itania for {W}ater {O}xidation},\ }\href
  {https://doi.org/10.1021/acs.jpcc.6b01481} {\bibfield  {journal} {\bibinfo
  {journal} {J. Phys. Chem. C}\ }\textbf {\bibinfo {volume} {120}},\ \bibinfo
  {pages} {6467} (\bibinfo {year} {2016})}\BibitemShut {NoStop}%
\bibitem [{\citenamefont {Eom}\ \emph {et~al.}(2015)\citenamefont {Eom},
  \citenamefont {Lim}, \citenamefont {Lee}, \citenamefont {Ryu},\ and\
  \citenamefont {Kwon}}]{Eom2015}%
  \BibitemOpen
  \bibfield  {author} {\bibinfo {author} {\bibfnamefont {J.-Y.}\ \bibnamefont
  {Eom}}, \bibinfo {author} {\bibfnamefont {S.-J.}\ \bibnamefont {Lim}},
  \bibinfo {author} {\bibfnamefont {S.-M.}\ \bibnamefont {Lee}}, \bibinfo
  {author} {\bibfnamefont {W.-H.}\ \bibnamefont {Ryu}},\ and\ \bibinfo {author}
  {\bibfnamefont {H.-S.}\ \bibnamefont {Kwon}},\ }\bibfield  {title} {\bibinfo
  {title} {Black titanium oxide nanoarray electrodes for high rate {L}i-ion
  microbatteries},\ }\href {https://doi.org/10.1039/C5TA01718A} {\bibfield
  {journal} {\bibinfo  {journal} {J. Mater. Chem. A}\ }\textbf {\bibinfo
  {volume} {3}},\ \bibinfo {pages} {11183} (\bibinfo {year}
  {2015})}\BibitemShut {NoStop}%
\bibitem [{\citenamefont {Gunnlaugsson}(2016)}]{Gunnlaugsson2016}%
  \BibitemOpen
  \bibfield  {author} {\bibinfo {author} {\bibfnamefont {H.~P.}\ \bibnamefont
  {Gunnlaugsson}},\ }\bibfield  {title} {\bibinfo {title} {Spreadsheet based
  analysis of {M}{\"o}ssbauer spectra},\ }\href
  {https://doi.org/10.1007/s10751-016-1271-z} {\bibfield  {journal} {\bibinfo
  {journal} {Hyperfine Interact.}\ }\textbf {\bibinfo {volume} {237}},\
  \bibinfo {pages} {79} (\bibinfo {year} {2016})}\BibitemShut {NoStop}%
\bibitem [{\citenamefont {Gunnlaugsson}(2006)}]{Gunnlaugsson2006}%
  \BibitemOpen
  \bibfield  {author} {\bibinfo {author} {\bibfnamefont {H.~P.}\ \bibnamefont
  {Gunnlaugsson}},\ }\bibfield  {title} {\bibinfo {title} {A simple model to
  extract hyperfine interaction distributions from {M}{\"o}ssbauer spectra},\
  }\href {https://doi.org/10.1007/s10751-006-9380-8} {\bibfield  {journal}
  {\bibinfo  {journal} {Hyperfine Interact.}\ }\textbf {\bibinfo {volume}
  {167}},\ \bibinfo {pages} {851} (\bibinfo {year} {2006})}\BibitemShut
  {NoStop}%
\bibitem [{\citenamefont {Umek}\ \emph {et~al.}(2014)\citenamefont {Umek},
  \citenamefont {Bittencourt}, \citenamefont {Guttmann}, \citenamefont
  {Gloter}, \citenamefont {\v{S}kapin},\ and\ \citenamefont
  {Ar\v{c}on}}]{Umek2014}%
  \BibitemOpen
  \bibfield  {author} {\bibinfo {author} {\bibfnamefont {P.}~\bibnamefont
  {Umek}}, \bibinfo {author} {\bibfnamefont {C.}~\bibnamefont {Bittencourt}},
  \bibinfo {author} {\bibfnamefont {P.}~\bibnamefont {Guttmann}}, \bibinfo
  {author} {\bibfnamefont {A.}~\bibnamefont {Gloter}}, \bibinfo {author}
  {\bibfnamefont {S.~D.}\ \bibnamefont {\v{S}kapin}},\ and\ \bibinfo {author}
  {\bibfnamefont {D.}~\bibnamefont {Ar\v{c}on}},\ }\bibfield  {title} {\bibinfo
  {title} {Mn\textsuperscript{2+} {S}ubstitutional {D}oping of
  {T}i{O}\textsubscript{2} {N}anoribbons: {A} {T}hree-{S}tep {A}pproach},\
  }\href {https://doi.org/10.1021/jp5063989} {\bibfield  {journal} {\bibinfo
  {journal} {J. Phys. Chem. C}\ }\textbf {\bibinfo {volume} {118}},\ \bibinfo
  {pages} {21250} (\bibinfo {year} {2014})}\BibitemShut {NoStop}%
\bibitem [{\citenamefont {Gunnlaugsson}\ \emph {et~al.}(2014)\citenamefont
  {Gunnlaugsson}, \citenamefont {Mantovan}, \citenamefont {Masenda},
  \citenamefont {M{\o}lholt}, \citenamefont {Johnston}, \citenamefont
  {Bharuth-Ram}, \citenamefont {Gislason}, \citenamefont {Langouche},
  \citenamefont {Naidoo}, \citenamefont {\'{O}lafsson}, \citenamefont {Svane},
  \citenamefont {Weyer},\ and\ \citenamefont {the
  ISOLDE~Collaboration}}]{Gunnlaugsson2014}%
  \BibitemOpen
  \bibfield  {author} {\bibinfo {author} {\bibfnamefont {H.~P.}\ \bibnamefont
  {Gunnlaugsson}}, \bibinfo {author} {\bibfnamefont {R.}~\bibnamefont
  {Mantovan}}, \bibinfo {author} {\bibfnamefont {H.}~\bibnamefont {Masenda}},
  \bibinfo {author} {\bibfnamefont {T.~E.}\ \bibnamefont {M{\o}lholt}},
  \bibinfo {author} {\bibfnamefont {K.}~\bibnamefont {Johnston}}, \bibinfo
  {author} {\bibfnamefont {K.}~\bibnamefont {Bharuth-Ram}}, \bibinfo {author}
  {\bibfnamefont {H.}~\bibnamefont {Gislason}}, \bibinfo {author}
  {\bibfnamefont {G.}~\bibnamefont {Langouche}}, \bibinfo {author}
  {\bibfnamefont {D.}~\bibnamefont {Naidoo}}, \bibinfo {author} {\bibfnamefont
  {S.}~\bibnamefont {\'{O}lafsson}}, \bibinfo {author} {\bibfnamefont
  {A.}~\bibnamefont {Svane}}, \bibinfo {author} {\bibfnamefont
  {G.}~\bibnamefont {Weyer}},\ and\ \bibinfo {author} {\bibnamefont {the
  ISOLDE~Collaboration}},\ }\bibfield  {title} {\bibinfo {title} {Defect
  annealing in {M}n/{F}e-implanted {T}i{O}$_2$ (rutile)},\ }\href
  {https://doi.org/doi:10.1088/0022-3727/47/6/065501} {\bibfield  {journal}
  {\bibinfo  {journal} {J. Phys. D: Appl. Phys.}\ }\textbf {\bibinfo {volume}
  {47}},\ \bibinfo {pages} {065501} (\bibinfo {year} {2014})}\BibitemShut
  {NoStop}%
\bibitem [{\citenamefont {M{\o}lholt}\ \emph {et~al.}(2010)\citenamefont
  {M{\o}lholt}, \citenamefont {Mantovan}, \citenamefont {Gunnlaugsson},
  \citenamefont {Naidoo}, \citenamefont {{\'O}lafsson}, \citenamefont
  {Bharuth-Ram}, \citenamefont {Fanciulli}, \citenamefont {Johnston},
  \citenamefont {Kobayashi}, \citenamefont {Langouche}, \citenamefont
  {Masenda}, \citenamefont {Sielemann}, \citenamefont {Weyer},\ and\
  \citenamefont {G{\'i}slason}}]{Molholt2010}%
  \BibitemOpen
  \bibfield  {author} {\bibinfo {author} {\bibfnamefont {T.~E.}\ \bibnamefont
  {M{\o}lholt}}, \bibinfo {author} {\bibfnamefont {R.}~\bibnamefont
  {Mantovan}}, \bibinfo {author} {\bibfnamefont {H.~P.}\ \bibnamefont
  {Gunnlaugsson}}, \bibinfo {author} {\bibfnamefont {D.}~\bibnamefont
  {Naidoo}}, \bibinfo {author} {\bibfnamefont {S.}~\bibnamefont
  {{\'O}lafsson}}, \bibinfo {author} {\bibfnamefont {K.}~\bibnamefont
  {Bharuth-Ram}}, \bibinfo {author} {\bibfnamefont {M.}~\bibnamefont
  {Fanciulli}}, \bibinfo {author} {\bibfnamefont {K.}~\bibnamefont {Johnston}},
  \bibinfo {author} {\bibfnamefont {Y.}~\bibnamefont {Kobayashi}}, \bibinfo
  {author} {\bibfnamefont {G.}~\bibnamefont {Langouche}}, \bibinfo {author}
  {\bibfnamefont {H.}~\bibnamefont {Masenda}}, \bibinfo {author} {\bibfnamefont
  {R.}~\bibnamefont {Sielemann}}, \bibinfo {author} {\bibfnamefont
  {G.}~\bibnamefont {Weyer}},\ and\ \bibinfo {author} {\bibfnamefont {H.~P.}\
  \bibnamefont {G{\'i}slason}},\ }\bibfield  {title} {\bibinfo {title}
  {Observation of spin-lattice relaxations of dilute fe$^{3+}$ in {M}g{O} by
  {M}{\"o}ssbauer spectroscopy},\ }\href
  {https://doi.org/10.1007/s10751-010-0214-3} {\bibfield  {journal} {\bibinfo
  {journal} {Hyperfine Interact.}\ }\textbf {\bibinfo {volume} {197}},\
  \bibinfo {pages} {89} (\bibinfo {year} {2010})}\BibitemShut {NoStop}%
\bibitem [{\citenamefont {Morgan}\ and\ \citenamefont
  {Watson}(2010)}]{Morgan2010}%
  \BibitemOpen
  \bibfield  {author} {\bibinfo {author} {\bibfnamefont {B.~J.}\ \bibnamefont
  {Morgan}}\ and\ \bibinfo {author} {\bibfnamefont {G.~W.}\ \bibnamefont
  {Watson}},\ }\bibfield  {title} {\bibinfo {title} {Intrinsic n-type {D}efect
  {F}ormation in {T}i{O}$_2$: {A} {C}omparison of {R}utile and {A}natase from
  {GGA+U} {C}alculations},\ }\href {https://doi.org/10.1021/jp9088047}
  {\bibfield  {journal} {\bibinfo  {journal} {J. Phys. Chem. C}\ }\textbf
  {\bibinfo {volume} {114}},\ \bibinfo {pages} {2321} (\bibinfo {year}
  {2010})}\BibitemShut {NoStop}%
\bibitem [{\citenamefont {Pan}\ \emph {et~al.}(2013)\citenamefont {Pan},
  \citenamefont {Yang}, \citenamefont {Fu}, \citenamefont {Zhang},\ and\
  \citenamefont {Xu}}]{Pan2013}%
  \BibitemOpen
  \bibfield  {author} {\bibinfo {author} {\bibfnamefont {X.}~\bibnamefont
  {Pan}}, \bibinfo {author} {\bibfnamefont {M.-Q.}\ \bibnamefont {Yang}},
  \bibinfo {author} {\bibfnamefont {X.}~\bibnamefont {Fu}}, \bibinfo {author}
  {\bibfnamefont {N.}~\bibnamefont {Zhang}},\ and\ \bibinfo {author}
  {\bibfnamefont {Y.-J.}\ \bibnamefont {Xu}},\ }\bibfield  {title} {\bibinfo
  {title} {Defective {T}i{O}$_2$ with oxygen vacancies: Synthesis, properties
  and photocatalytic applications},\ }\href
  {https://doi.org/10.1039/C3NR00476G} {\bibfield  {journal} {\bibinfo
  {journal} {Nanoscale}\ }\textbf {\bibinfo {volume} {5}},\ \bibinfo {pages}
  {3601} (\bibinfo {year} {2013})}\BibitemShut {NoStop}%
\bibitem [{\citenamefont {Robinson}\ \emph {et~al.}(2014)\citenamefont
  {Robinson}, \citenamefont {Marks},\ and\ \citenamefont
  {Lumpkin}}]{Robinson2014}%
  \BibitemOpen
  \bibfield  {author} {\bibinfo {author} {\bibfnamefont {M.}~\bibnamefont
  {Robinson}}, \bibinfo {author} {\bibfnamefont {N.~A.}\ \bibnamefont
  {Marks}},\ and\ \bibinfo {author} {\bibfnamefont {G.~R.}\ \bibnamefont
  {Lumpkin}},\ }\bibfield  {title} {\bibinfo {title} {Structural dependence of
  threshold displacement energies in rutile, anatase and brookite
  {T}i{O}$_2$},\ }\href {https://doi.org/10.1016/j.matchemphys.2014.05.006}
  {\bibfield  {journal} {\bibinfo  {journal} {Mater. Chem. Phys.}\ }\textbf
  {\bibinfo {volume} {147}},\ \bibinfo {pages} {311} (\bibinfo {year}
  {2014})}\BibitemShut {NoStop}%
\bibitem [{\citenamefont {Ghicov}\ \emph {et~al.}(2006)\citenamefont {Ghicov},
  \citenamefont {Macak}, \citenamefont {Tsuchiya}, \citenamefont {Kunze},
  \citenamefont {Haeublein}, \citenamefont {Frey},\ and\ \citenamefont
  {Schmuki}}]{Ghicov2006}%
  \BibitemOpen
  \bibfield  {author} {\bibinfo {author} {\bibfnamefont {A.}~\bibnamefont
  {Ghicov}}, \bibinfo {author} {\bibfnamefont {J.~M.}\ \bibnamefont {Macak}},
  \bibinfo {author} {\bibfnamefont {H.}~\bibnamefont {Tsuchiya}}, \bibinfo
  {author} {\bibfnamefont {J.}~\bibnamefont {Kunze}}, \bibinfo {author}
  {\bibfnamefont {V.}~\bibnamefont {Haeublein}}, \bibinfo {author}
  {\bibfnamefont {L.}~\bibnamefont {Frey}},\ and\ \bibinfo {author}
  {\bibfnamefont {P.}~\bibnamefont {Schmuki}},\ }\bibfield  {title} {\bibinfo
  {title} {Ion {I}mplantation and {A}nnealing for an {E}fficient {N}-{D}oping
  of {T}i{O}$_2$ {N}anotubes},\ }\href {https://doi.org/10.1021/nl0600979}
  {\bibfield  {journal} {\bibinfo  {journal} {Nano Lett.}\ }\textbf {\bibinfo
  {volume} {6}},\ \bibinfo {pages} {1080} (\bibinfo {year} {2006})}\BibitemShut
  {NoStop}%
\bibitem [{\citenamefont {De\'ak}\ \emph {et~al.}(2011)\citenamefont {De\'ak},
  \citenamefont {Aradi},\ and\ \citenamefont {Frauenheim}}]{Deak2011}%
  \BibitemOpen
  \bibfield  {author} {\bibinfo {author} {\bibfnamefont {P.}~\bibnamefont
  {De\'ak}}, \bibinfo {author} {\bibfnamefont {B.}~\bibnamefont {Aradi}},\ and\
  \bibinfo {author} {\bibfnamefont {T.}~\bibnamefont {Frauenheim}},\ }\bibfield
   {title} {\bibinfo {title} {Polaronic effects in {T}i{O}${}_{2}$ calculated
  by the {HSE06} hybrid functional: {D}opant passivation by carrier
  self-trapping},\ }\href {https://doi.org/10.1103/PhysRevB.83.155207}
  {\bibfield  {journal} {\bibinfo  {journal} {Phys. Rev. B}\ }\textbf {\bibinfo
  {volume} {83}},\ \bibinfo {pages} {155207} (\bibinfo {year}
  {2011})}\BibitemShut {NoStop}%
\bibitem [{\citenamefont {Dolci}\ \emph {et~al.}(2007)\citenamefont {Dolci},
  \citenamefont {Chio}, \citenamefont {Baricco},\ and\ \citenamefont
  {Giamello}}]{Dolci2007}%
  \BibitemOpen
  \bibfield  {author} {\bibinfo {author} {\bibfnamefont {F.}~\bibnamefont
  {Dolci}}, \bibinfo {author} {\bibfnamefont {M.~D.}\ \bibnamefont {Chio}},
  \bibinfo {author} {\bibfnamefont {M.}~\bibnamefont {Baricco}},\ and\ \bibinfo
  {author} {\bibfnamefont {E.}~\bibnamefont {Giamello}},\ }\bibfield  {title}
  {\bibinfo {title} {Niobium pentoxide as promoter in the mixed
  {M}g{H}$_2$/{N}b{2}o{5} system for hydrogen storage: a multitechnique
  investigation of the {H}$_2$ uptake},\ }\href
  {https://doi.org/10.1007/s10853-007-1567-0} {\bibfield  {journal} {\bibinfo
  {journal} {J. Mater. Sci.}\ }\textbf {\bibinfo {volume} {42}},\ \bibinfo
  {pages} {7180} (\bibinfo {year} {2007})}\BibitemShut {NoStop}%
\bibitem [{\citenamefont {Chen}\ \emph {et~al.}(2013)\citenamefont {Chen},
  \citenamefont {He}, \citenamefont {Wang}, \citenamefont {Chan},\ and\
  \citenamefont {Yan}}]{ChenWan2013}%
  \BibitemOpen
  \bibfield  {author} {\bibinfo {author} {\bibfnamefont {W.~P.}\ \bibnamefont
  {Chen}}, \bibinfo {author} {\bibfnamefont {K.~F.}\ \bibnamefont {He}},
  \bibinfo {author} {\bibfnamefont {Y.}~\bibnamefont {Wang}}, \bibinfo {author}
  {\bibfnamefont {H.~L.~W.}\ \bibnamefont {Chan}},\ and\ \bibinfo {author}
  {\bibfnamefont {Z.}~\bibnamefont {Yan}},\ }\bibfield  {title} {\bibinfo
  {title} {Highly mobile and reactive state of hydrogen in metal oxide
  semiconductors at room temperature},\ }\href@noop {} {\bibfield  {journal}
  {\bibinfo  {journal} {Sci. Rep.}\ }\textbf {\bibinfo {volume} {3}},\ \bibinfo
  {pages} {3149 EP } (\bibinfo {year} {2013})},\ \bibinfo {note}
  {article}\BibitemShut {NoStop}%
\bibitem [{\citenamefont {Chen}\ \emph {et~al.}(2008)\citenamefont {Chen},
  \citenamefont {Wang},\ and\ \citenamefont {Chan}}]{ChenWan2008}%
  \BibitemOpen
  \bibfield  {author} {\bibinfo {author} {\bibfnamefont {W.~P.}\ \bibnamefont
  {Chen}}, \bibinfo {author} {\bibfnamefont {Y.}~\bibnamefont {Wang}},\ and\
  \bibinfo {author} {\bibfnamefont {H.~L.~W.}\ \bibnamefont {Chan}},\
  }\bibfield  {title} {\bibinfo {title} {Hydrogen: {A} metastable donor in
  {T}i{O}$_2$ single crystals},\ }\href {https://doi.org/10.1063/1.2900957}
  {\bibfield  {journal} {\bibinfo  {journal} {Appl. Phys. Lett.}\ }\textbf
  {\bibinfo {volume} {92}},\ \bibinfo {pages} {112907} (\bibinfo {year}
  {2008})}\BibitemShut {NoStop}%
\bibitem [{\citenamefont {Kobayashi}\ \emph {et~al.}(2012)\citenamefont
  {Kobayashi}, \citenamefont {Hernandez}, \citenamefont {Sakaguchi},
  \citenamefont {Yajima}, \citenamefont {Roisnel}, \citenamefont {Tsujimoto},
  \citenamefont {Morita}, \citenamefont {Noda}, \citenamefont {Mogami},
  \citenamefont {Kitada}, \citenamefont {Ohkura}, \citenamefont {Hosokawa},
  \citenamefont {Li}, \citenamefont {Hayashi}, \citenamefont {Kusano},
  \citenamefont {Kim}, \citenamefont {Tsuji}, \citenamefont {Fujiwara},
  \citenamefont {Matsushita}, \citenamefont {Yoshimura}, \citenamefont
  {Takegoshi}, \citenamefont {Inoue}, \citenamefont {Takano},\ and\
  \citenamefont {Kageyama}}]{Kobayashi2012}%
  \BibitemOpen
  \bibfield  {author} {\bibinfo {author} {\bibfnamefont {Y.}~\bibnamefont
  {Kobayashi}}, \bibinfo {author} {\bibfnamefont {O.~J.}\ \bibnamefont
  {Hernandez}}, \bibinfo {author} {\bibfnamefont {T.}~\bibnamefont
  {Sakaguchi}}, \bibinfo {author} {\bibfnamefont {T.}~\bibnamefont {Yajima}},
  \bibinfo {author} {\bibfnamefont {T.}~\bibnamefont {Roisnel}}, \bibinfo
  {author} {\bibfnamefont {Y.}~\bibnamefont {Tsujimoto}}, \bibinfo {author}
  {\bibfnamefont {M.}~\bibnamefont {Morita}}, \bibinfo {author} {\bibfnamefont
  {Y.}~\bibnamefont {Noda}}, \bibinfo {author} {\bibfnamefont {Y.}~\bibnamefont
  {Mogami}}, \bibinfo {author} {\bibfnamefont {A.}~\bibnamefont {Kitada}},
  \bibinfo {author} {\bibfnamefont {M.}~\bibnamefont {Ohkura}}, \bibinfo
  {author} {\bibfnamefont {S.}~\bibnamefont {Hosokawa}}, \bibinfo {author}
  {\bibfnamefont {Z.}~\bibnamefont {Li}}, \bibinfo {author} {\bibfnamefont
  {K.}~\bibnamefont {Hayashi}}, \bibinfo {author} {\bibfnamefont
  {Y.}~\bibnamefont {Kusano}}, \bibinfo {author} {\bibfnamefont {J.~e.}\
  \bibnamefont {Kim}}, \bibinfo {author} {\bibfnamefont {N.}~\bibnamefont
  {Tsuji}}, \bibinfo {author} {\bibfnamefont {A.}~\bibnamefont {Fujiwara}},
  \bibinfo {author} {\bibfnamefont {Y.}~\bibnamefont {Matsushita}}, \bibinfo
  {author} {\bibfnamefont {K.}~\bibnamefont {Yoshimura}}, \bibinfo {author}
  {\bibfnamefont {K.}~\bibnamefont {Takegoshi}}, \bibinfo {author}
  {\bibfnamefont {M.}~\bibnamefont {Inoue}}, \bibinfo {author} {\bibfnamefont
  {M.}~\bibnamefont {Takano}},\ and\ \bibinfo {author} {\bibfnamefont
  {H.}~\bibnamefont {Kageyama}},\ }\bibfield  {title} {\bibinfo {title} {An
  oxyhydride of batio$_3$ exhibiting hydride exchange and electronic
  conductivity},\ }\href {https://doi.org/10.1038/nmat3302} {\bibfield
  {journal} {\bibinfo  {journal} {Nat. Mater.}\ }\textbf {\bibinfo {volume}
  {11}},\ \bibinfo {pages} {507 EP } (\bibinfo {year} {2012})}\BibitemShut
  {NoStop}%
\bibitem [{\citenamefont {Hupfer}\ \emph {et~al.}(2017)\citenamefont {Hupfer},
  \citenamefont {Monakhov}, \citenamefont {Svensson}, \citenamefont
  {Chaplygin},\ and\ \citenamefont {Lavrov}}]{Hupfer2017}%
  \BibitemOpen
  \bibfield  {author} {\bibinfo {author} {\bibfnamefont {A.~J.}\ \bibnamefont
  {Hupfer}}, \bibinfo {author} {\bibfnamefont {E.~V.}\ \bibnamefont
  {Monakhov}}, \bibinfo {author} {\bibfnamefont {B.~G.}\ \bibnamefont
  {Svensson}}, \bibinfo {author} {\bibfnamefont {I.}~\bibnamefont
  {Chaplygin}},\ and\ \bibinfo {author} {\bibfnamefont {E.~V.}\ \bibnamefont
  {Lavrov}},\ }\bibfield  {title} {\bibinfo {title} {Hydrogen motion in rutile
  {T}i{O}$_2$},\ }\href {https://doi.org/10.1038/s41598-017-16660-3} {\bibfield
   {journal} {\bibinfo  {journal} {Sci. Rep.}\ }\textbf {\bibinfo {volume}
  {7}},\ \bibinfo {pages} {17065} (\bibinfo {year} {2017})}\BibitemShut
  {NoStop}%
\bibitem [{\citenamefont {Sheikholeslam}\ \emph {et~al.}(2016)\citenamefont
  {Sheikholeslam}, \citenamefont {Manzano}, \citenamefont {Grecu},\ and\
  \citenamefont {Ivanov}}]{Sheikholeslam2016}%
  \BibitemOpen
  \bibfield  {author} {\bibinfo {author} {\bibfnamefont {S.~A.}\ \bibnamefont
  {Sheikholeslam}}, \bibinfo {author} {\bibfnamefont {H.}~\bibnamefont
  {Manzano}}, \bibinfo {author} {\bibfnamefont {C.}~\bibnamefont {Grecu}},\
  and\ \bibinfo {author} {\bibfnamefont {A.}~\bibnamefont {Ivanov}},\
  }\bibfield  {title} {\bibinfo {title} {Reduced hydrogen diffusion in strained
  amorphous {S}i{O}$_2$: Understanding ageing in {MOSFET} devices},\ }\href
  {https://doi.org/10.1039/C6TC02647H} {\bibfield  {journal} {\bibinfo
  {journal} {J. Mater. Chem. C}\ }\textbf {\bibinfo {volume} {4}},\ \bibinfo
  {pages} {8104} (\bibinfo {year} {2016})}\BibitemShut {NoStop}%
\bibitem [{\citenamefont {Chaplygin}\ \emph {et~al.}(2018)\citenamefont
  {Chaplygin}, \citenamefont {Herklotz},\ and\ \citenamefont
  {Lavrov}}]{Chaplygin2018}%
  \BibitemOpen
  \bibfield  {author} {\bibinfo {author} {\bibfnamefont {I.}~\bibnamefont
  {Chaplygin}}, \bibinfo {author} {\bibfnamefont {F.}~\bibnamefont
  {Herklotz}},\ and\ \bibinfo {author} {\bibfnamefont {E.~V.}\ \bibnamefont
  {Lavrov}},\ }\bibfield  {title} {\bibinfo {title} {Reorientation kinetics of
  hydroxyl groups in anatase {T}i{O}$_2$},\ }\href
  {https://doi.org/10.1063/1.5039584} {\bibfield  {journal} {\bibinfo
  {journal} {J. Chem. Phys.}\ }\textbf {\bibinfo {volume} {149}},\ \bibinfo
  {pages} {044507} (\bibinfo {year} {2018})}\BibitemShut {NoStop}%
\bibitem [{\citenamefont {Nandasiri}\ \emph {et~al.}(2015)\citenamefont
  {Nandasiri}, \citenamefont {Shutthanandan}, \citenamefont {Manandhar},
  \citenamefont {Schwarz}, \citenamefont {Oxenford}, \citenamefont {Kennedy},
  \citenamefont {Thevuthasan},\ and\ \citenamefont
  {Henderson}}]{Nandasiri2015}%
  \BibitemOpen
  \bibfield  {author} {\bibinfo {author} {\bibfnamefont {M.~I.}\ \bibnamefont
  {Nandasiri}}, \bibinfo {author} {\bibfnamefont {V.}~\bibnamefont
  {Shutthanandan}}, \bibinfo {author} {\bibfnamefont {S.}~\bibnamefont
  {Manandhar}}, \bibinfo {author} {\bibfnamefont {A.~M.}\ \bibnamefont
  {Schwarz}}, \bibinfo {author} {\bibfnamefont {L.}~\bibnamefont {Oxenford}},
  \bibinfo {author} {\bibfnamefont {J.~V.}\ \bibnamefont {Kennedy}}, \bibinfo
  {author} {\bibfnamefont {S.}~\bibnamefont {Thevuthasan}},\ and\ \bibinfo
  {author} {\bibfnamefont {M.~A.}\ \bibnamefont {Henderson}},\ }\bibfield
  {title} {\bibinfo {title} {Instability of {H}ydrogenated {T}i{O}$_2$},\
  }\href {https://doi.org/10.1021/acs.jpclett.5b02219} {\bibfield  {journal}
  {\bibinfo  {journal} {J. Phys. Chem. Lett.}\ }\textbf {\bibinfo {volume}
  {6}},\ \bibinfo {pages} {4627} (\bibinfo {year} {2015})}\BibitemShut
  {NoStop}%
\bibitem [{\citenamefont {Johnson}\ \emph {et~al.}(1975)\citenamefont
  {Johnson}, \citenamefont {Paek},\ and\ \citenamefont {DeFord}}]{Johnson1975}%
  \BibitemOpen
  \bibfield  {author} {\bibinfo {author} {\bibfnamefont {O.~W.}\ \bibnamefont
  {Johnson}}, \bibinfo {author} {\bibfnamefont {S.-H.}\ \bibnamefont {Paek}},\
  and\ \bibinfo {author} {\bibfnamefont {J.~W.}\ \bibnamefont {DeFord}},\
  }\bibfield  {title} {\bibinfo {title} {Diffusion of {H} and {D} in tio$_2$:
  {S}uppression of internal fields by isotope exchange},\ }\href
  {https://doi.org/10.1063/1.322206} {\bibfield  {journal} {\bibinfo  {journal}
  {J. Appl. Phys.}\ }\textbf {\bibinfo {volume} {46}},\ \bibinfo {pages} {1026}
  (\bibinfo {year} {1975})}\BibitemShut {NoStop}%
\bibitem [{\citenamefont {Koch}\ \emph {et~al.}(2014)\citenamefont {Koch},
  \citenamefont {Lavrov},\ and\ \citenamefont {Weber}}]{Koch2014}%
  \BibitemOpen
  \bibfield  {author} {\bibinfo {author} {\bibfnamefont {S.~G.}\ \bibnamefont
  {Koch}}, \bibinfo {author} {\bibfnamefont {E.~V.}\ \bibnamefont {Lavrov}},\
  and\ \bibinfo {author} {\bibfnamefont {J.}~\bibnamefont {Weber}},\ }\bibfield
   {title} {\bibinfo {title} {Towards understanding the hydrogen molecule in
  {Z}no},\ }\href {https://doi.org/10.1103/PhysRevB.90.205212} {\bibfield
  {journal} {\bibinfo  {journal} {Phys. Rev. B}\ }\textbf {\bibinfo {volume}
  {90}},\ \bibinfo {pages} {205212} (\bibinfo {year} {2014})}\BibitemShut
  {NoStop}%
\bibitem [{\citenamefont {Henderson}(1995)}]{Henderson1995}%
  \BibitemOpen
  \bibfield  {author} {\bibinfo {author} {\bibfnamefont {M.~A.}\ \bibnamefont
  {Henderson}},\ }\bibfield  {title} {\bibinfo {title} {Mechanism for the
  bulk-assisted reoxidation of ion sputtered {T}i{O}$_2$ surfaces: Diffusion of
  oxygen to the surface or titanium to the bulk?},\ }\href
  {https://doi.org/10.1016/0039-6028(95)00849-7} {\bibfield  {journal}
  {\bibinfo  {journal} {Surf. Sci.}\ }\textbf {\bibinfo {volume} {343}},\
  \bibinfo {pages} {L1156 } (\bibinfo {year} {1995})}\BibitemShut {NoStop}%
\bibitem [{\citenamefont {Rodr\'{i}guez-Torres}\ \emph
  {et~al.}(2008)\citenamefont {Rodr\'{i}guez-Torres}, \citenamefont {Cabrera},
  \citenamefont {Errico}, \citenamefont {Ad\'{a}n}, \citenamefont {Requejo},
  \citenamefont {Weissmann},\ and\ \citenamefont {Stewart}}]{Rod2008}%
  \BibitemOpen
  \bibfield  {author} {\bibinfo {author} {\bibfnamefont {C.~E.}\ \bibnamefont
  {Rodr\'{i}guez-Torres}}, \bibinfo {author} {\bibfnamefont {A.~F.}\
  \bibnamefont {Cabrera}}, \bibinfo {author} {\bibfnamefont {L.~A.}\
  \bibnamefont {Errico}}, \bibinfo {author} {\bibfnamefont {C.}~\bibnamefont
  {Ad\'{a}n}}, \bibinfo {author} {\bibfnamefont {F.~G.}\ \bibnamefont
  {Requejo}}, \bibinfo {author} {\bibfnamefont {M.}~\bibnamefont {Weissmann}},\
  and\ \bibinfo {author} {\bibfnamefont {S.~J.}\ \bibnamefont {Stewart}},\
  }\bibfield  {title} {\bibinfo {title} {Local structure and magnetic behaviour
  of {F}e-doped {T}i{O}$_2$ anatase nanoparticles: Experiments and
  calculations},\ }\href@noop {} {\bibfield  {journal} {\bibinfo  {journal} {J.
  Phys. Condens. Matter}\ }\textbf {\bibinfo {volume} {20}},\ \bibinfo {pages}
  {135210} (\bibinfo {year} {2008})}\BibitemShut {NoStop}%
\bibitem [{\citenamefont {Kordatos}\ \emph {et~al.}(2018)\citenamefont
  {Kordatos}, \citenamefont {Kelaidis},\ and\ \citenamefont
  {Chroneos}}]{KORDATOS2018}%
  \BibitemOpen
  \bibfield  {author} {\bibinfo {author} {\bibfnamefont {A.}~\bibnamefont
  {Kordatos}}, \bibinfo {author} {\bibfnamefont {N.}~\bibnamefont {Kelaidis}},\
  and\ \bibinfo {author} {\bibfnamefont {A.}~\bibnamefont {Chroneos}},\
  }\bibfield  {title} {\bibinfo {title} {Migration of sodium and lithium
  interstitials in anatase {T}i{O}\textsubscript{2}},\ }\href
  {https://doi.org/10.1016/j.ssi.2017.12.003} {\bibfield  {journal} {\bibinfo
  {journal} {Solid State Ionics}\ }\textbf {\bibinfo {volume} {315}},\ \bibinfo
  {pages} {40 } (\bibinfo {year} {2018})}\BibitemShut {NoStop}%
\bibitem [{\citenamefont {Chen}\ and\ \citenamefont {Yang}(2007)}]{ChenYa2007}%
  \BibitemOpen
  \bibfield  {author} {\bibinfo {author} {\bibfnamefont {Y.}~\bibnamefont
  {Chen}}\ and\ \bibinfo {author} {\bibfnamefont {D.}~\bibnamefont {Yang}},\
  }\href {https://doi.org/10.1002/9783527611423.ch2} {\emph {\bibinfo {title}
  {M{\"o}ssbauer {E}ffect in {L}attice {D}ynamics}}}\ (\bibinfo  {publisher}
  {Wiley-Blackwell},\ \bibinfo {year} {2007})\ Chap.~\bibinfo {chapter} {2},
  pp.\ \bibinfo {pages} {29--77}\BibitemShut {NoStop}%
\bibitem [{\citenamefont {Gunnlaugsson}\ and\ \citenamefont
  {Masenda}(2019)}]{Gunnlaugsson2019}%
  \BibitemOpen
  \bibfield  {author} {\bibinfo {author} {\bibfnamefont {H.}~\bibnamefont
  {Gunnlaugsson}}\ and\ \bibinfo {author} {\bibfnamefont {H.}~\bibnamefont
  {Masenda}},\ }\bibfield  {title} {\bibinfo {title} {M{\"o}ssbauer
  isomer-shift of ferrous iron impurities in ionic and covalent binary
  compounds},\ }\href {https://doi.org/10.1016/j.jpcs.2018.12.037} {\bibfield
  {journal} {\bibinfo  {journal} {J. Phys. Chem. Solids}\ }\textbf {\bibinfo
  {volume} {129}},\ \bibinfo {pages} {151 } (\bibinfo {year}
  {2019})}\BibitemShut {NoStop}%
\bibitem [{\citenamefont {Zhu}\ \emph {et~al.}(2007)\citenamefont {Zhu},
  \citenamefont {Liu}, \citenamefont {Wei}, \citenamefont {Fan},\ and\
  \citenamefont {Li}}]{Sanyuan2007}%
  \BibitemOpen
  \bibfield  {author} {\bibinfo {author} {\bibfnamefont {S.}~\bibnamefont
  {Zhu}}, \bibinfo {author} {\bibfnamefont {W.}~\bibnamefont {Liu}}, \bibinfo
  {author} {\bibfnamefont {S.}~\bibnamefont {Wei}}, \bibinfo {author}
  {\bibfnamefont {C.}~\bibnamefont {Fan}},\ and\ \bibinfo {author}
  {\bibfnamefont {Y.}~\bibnamefont {Li}},\ }\bibfield  {title} {\bibinfo
  {title} {Local structure around {I}ron {I}ons in {A}natase {T}i{O}$_2$},\
  }\href {https://doi.org/10.1063/1.2644492} {\bibfield  {journal} {\bibinfo
  {journal} {AIP Conf. Proc.}\ }\textbf {\bibinfo {volume} {882}},\ \bibinfo
  {pages} {253} (\bibinfo {year} {2007})}\BibitemShut {NoStop}%
\end{thebibliography}%


%

\end{document}